\documentclass[amsmath,amssymb,superscriptaddress,nobalancelastpage,prb,twocolumn]{revtex4-1}
\usepackage{color}
\usepackage[dvips]{graphicx}

\newcommand{\LSCO}{La$_{2-x}$Sr$_x$CuO$_4$}

\begin{document}

\title{Spin density wave induced disordering of the vortex lattice in superconducting
La$_{2-x}$Sr$_x$CuO$_4$}
\author{J.\ Chang}

\address{Laboratory for Neutron Scattering, Paul Scherrer Institut, CH-5232 Villigen, Switzerland}
\address{Institut de la materi\`ere complexe, Ecole Polytechnique Fed\'ed\'erale de Lausanne (EPFL), CH-1015 Lausanne, Switzerland}
\author{J.\,S.\ White}
\address{Laboratory for Neutron Scattering, Paul Scherrer Institut, CH-5232 Villigen, Switzerland}
\address{Institut de la materi\`ere complexe, Ecole Polytechnique Fed\'ed\'erale de Lausanne (EPFL), CH-1015 Lausanne, Switzerland}
\author{M.\ Laver}
\address{Laboratory for Neutron Scattering, Paul Scherrer Institut, CH-5232 Villigen, Switzerland}
\address{Department of Physics, Technical University of Denmark, DK-2800 Kgs.\ Lyngby, Denmark}
\address{Nano-Science Center, Niels Bohr Institute, University of Copenhagen, DK-2100 K\o benhavn, Denmark}
\author{C.\,J.\ Bowell}
\address{School of Physics and Astronomy, The University of Birmingham, Birmingham B15 2TT, United Kingdom}
\author{S.\,P.\ Brown}
\address{School of Physics and Astronomy, The University of Birmingham, Birmingham B15 2TT, United Kingdom}
\author{A.\,T.\ Holmes}
\address{School of Physics and Astronomy, The University of Birmingham, Birmingham B15 2TT, United Kingdom}
\author{L.\ Maechler}
\address{Laboratory for Neutron Scattering, Paul Scherrer Institut, CH-5232 Villigen, Switzerland}
\author{S.\ Str\"{a}ssle}
\address{Physik-Institut der Universitat Z\"{u}rich, CH-8057 Z\"{u}rich, Switzerland }
\author{R.\ Gilardi}
\address{Laboratory for Neutron Scattering, Paul Scherrer Institut, CH-5232 Villigen, Switzerland}
\author{S.\ Gerber}
\address{Laboratory for Neutron Scattering, Paul Scherrer Institut, CH-5232 Villigen, Switzerland}
\author{T.\ Kurosawa }
\affiliation{Department of Physics, Hokkaido University - Sapporo 060-0810, Japan}
\author{N.\ Momono}
\affiliation{Department of Physics, Hokkaido University - Sapporo 060-0810, Japan}
\author{M.\ Oda}
\affiliation{Department of Physics, Hokkaido University - Sapporo 060-0810, Japan}
\author{M.\ Ido}
\affiliation{Department of Physics, Hokkaido University - Sapporo 060-0810, Japan}
\author{O.\,J.\ Lipscombe}
\affiliation{H.\ H.\ Wills Physics Laboratory, University of Bristol, Bristol, BS8 1TL, United Kingdom}
\author{S. M.\ Hayden}
\affiliation{H.\ H.\ Wills Physics Laboratory, University of Bristol, Bristol, BS8 1TL, United Kingdom}
\author{C.\,D.\ Dewhurst}
\address{Institut Laue-Langevin, 6 rue Jules Horowitz, 38042 Grenoble, France}
\author{R.\ Vavrin}
\address{Laboratory for Neutron Scattering, Paul Scherrer Institut, CH-5232 Villigen, Switzerland}
\author{J.\ Gavilano}
\address{Laboratory for Neutron Scattering, Paul Scherrer Institut, CH-5232 Villigen, Switzerland}
\author{J.\ Kohlbrecher}
\address{Laboratory for Neutron Scattering, Paul Scherrer Institut, CH-5232 Villigen, Switzerland}
\author{E.\,M.\ Forgan}
\address{School of Physics and Astronomy, The University of Birmingham, Birmingham B15 2TT, United Kingdom}
\author{J.\ Mesot}
\address{Laboratory for Neutron Scattering, Paul Scherrer Institut, CH-5232 Villigen, Switzerland}
\address{Institut de la materi\`ere complexe, Ecole Polytechnique Fed\'ed\'erale de Lausanne (EPFL), CH-1015 Lausanne, Switzerland}

\date{\today}

\begin{abstract}
 We use  small angle neutron scattering to study the superconducting vortex
lattice  in La$_{2-x}$Sr$_x$CuO$_4$ as a function of doping and magnetic field.
We show that near optimally doping
 the vortex lattice coordination and the superconducting coherence length $\xi$
 are  controlled by a van-Hove singularity crossing the Fermi level
 near the Brillouin zone boundary.
 The vortex lattice properties change dramatically as a spin-density-wave instability
is approached upon underdoping.
 The Bragg glass paradigm provides a good description of this regime
and suggests that SDW order acts as a novel source of disorder on the vortex lattice.
\end{abstract}

\maketitle

\section{Introduction}
A commonality across the
borocarbides,\citep{Can98,Mul01,Gup06,Bud06} cuprates,\citep{Lak02}
 ferro-pnictides,\citep{Esk11} heavy-fermion\citep{Mon07}
and organic superconductors\citep{Jer83}
is the coexistence of magnetism and superconductivity. The corresponding order parameters
typically compete\citep{Voj09} and often a small perturbation is sufficient to tip the balance
between the two.
For example, the magnetism carried by the rare earth ions R in the borocarbides RNi$_2$B$_2$C can lead
to nearly reentrant superconducting phase diagrams, and spontaneously forming
superconducting vortices at zero applied field.\citep{Can98,Mul01,Gup06,Bud06}
Equally, vortices induced under applied fields may permit enhanced
magnetic correlations in the core regions where the superconducting order parameter is
suppressed.\citep{Dem01,Hof02,Sac03,And09}
This idea was put forward to explain  field-induced and -enhanced magnetic correlations
observed in the cuprate superconductor \LSCO\ (LSCO).\citep{Lak01,Lak02}

Although the effect of static magnetism on moving vortices was recently
considered theoretically,\citep{She11}
little is known about how the presence of magnetic correlations affects the
arrangement of vortices.
Here we address the problem from an experimental point of view.
When magnetism and superconductivity coexist
there are at least four relevant length scales:
the penetration depth $\lambda$, the vortex core size $\xi$, the vortex spacing $a_0$, and the
magnetic correlation length $\zeta$.
The  vortex spacing $a_0\propto H^{-0.5}$ scales with
the applied magnetic field $\mu_0 H$ and in LSCO the magnetic correlation
length can be tuned by varying the doping concentration.
Using small angle
neutron scattering (SANS) we have studied two different regimes (see Fig.~1);
(i) far away from the magnetic ordering where $\xi,\zeta\ll a_0$
and (ii) entering  the magnetic phase where
$\zeta\sim a_0$. In the first regime, where static magnetism is absent,
the vortex lattice (VL) structure and core size are understood from
pure fermiological considerations.
In the second regime with static long-range magnetism, the vortex arrangement
exhibits increasing disorder. We find these regimes to be well-described
within the topical `Bragg glass' paradigm,
where disorder results in an algebraic decay of the translational order of the vortices.\citep{Gia95,Bog01}
VL disordering is usually driven by effects extrinsic to superconductivity
such as rare earth magnetism in RNi$_2$B$_2$C,\citep{Can98,Mul01,Gup06,Bud06,Vin05,Blu06,Mol11}
or sample impurities and crystalline defects.\citep{Lav08}
In contrast, magnetic and SC order parameters are intertwined in LSCO;
we show that this provides a novel and tunable source of VL disorder.

\begin{figure}[h!]
 \begin{center}
\includegraphics[width=0.46\textwidth]{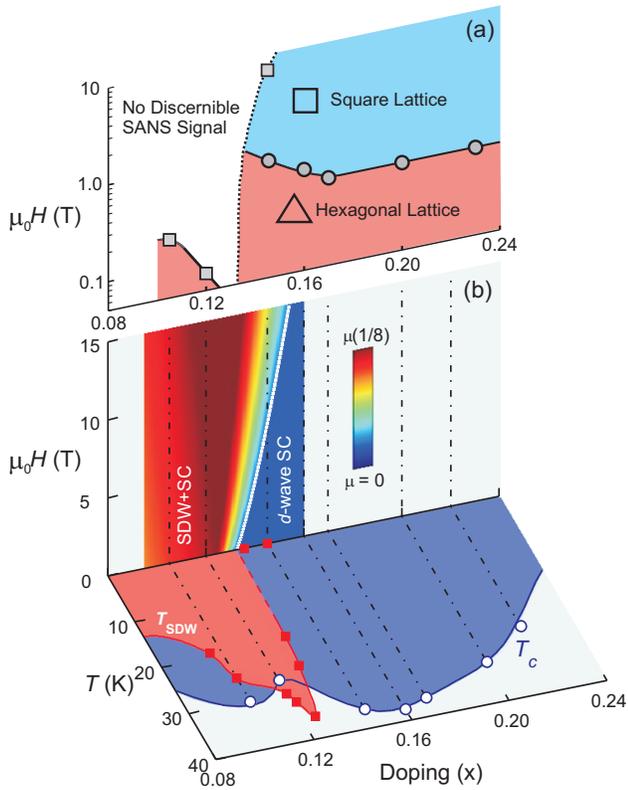}
\caption{ (Color online) Phase diagram of (a) the ($T<3$~K) vortex lattice structure and (b) the magnetism
in La$_{2-x}$Sr$_x$CuO$_4$, both revealed by neutron diffraction.
Data points in (a) for $x=0.12$, 0.145, 0.16 and 0.22 are from this work
whereas $x=0.105$, 0.17, and 0.20 are
from Ref.~\onlinecite{Div04,Gil02,Gil04}.
Circular points are defined by the onset of a square VL coordination.
The field-doping plane in (b), adapted from Ref.~\onlinecite{Cha08},  shows
schematically the ordered SDW-moment normalized to that at the 1/8-doping.
Field-induced order was first reported by Khaykovich \textit{et al.}\citep{Kha05} and later
confirmed in Ref.~\onlinecite{Cha08}.
The temperature-doping plane shows the superconducting dome together
with the onset  of static incommensurate SDW order $T_{SDW}$ as seen
by neutron diffraction.\citep{Cha08,Kof09} We remark that
a similar phase diagram was proposed for YBa$_2$Cu$_3$O$_y$.\citep{Hau10}
The dashed lines indicate the samples studied in this paper.
}\label{fig:fig1a}
  \end{center}
\end{figure}

\begin{figure}[h!]
 \begin{center}
\includegraphics[width=0.46\textwidth]{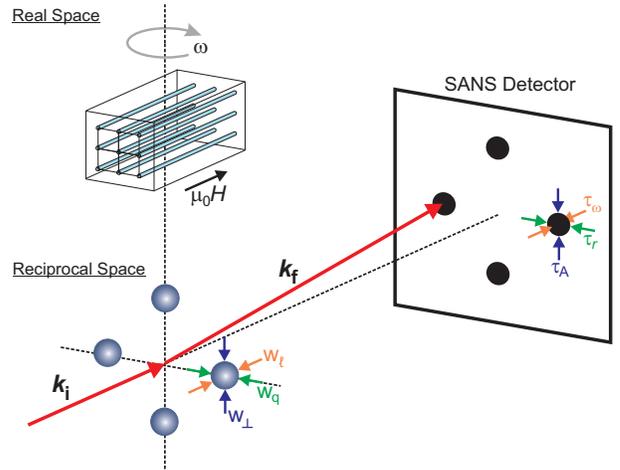}
\caption{ A schematic diagram illustrating the experimental
geometry chosen for our experiments. A square VL in real space
forms a two-dimensional reciprocal VL, the properties of which
are recorded by SANS using a position-sensitive detector.
The Bragg spots in reciprocal space exhibit finite
widths $w_{\ell}$, $w_{\rm q}$ and $w_{\perp}$ that are
dependent on both the instrumental resolution and properties of
the VL. These three lengths are estimated at the detector by
 recording the angular widths $\tau_{\rm r}$ and $\tau_{\rm A}$ within
 the detector plane, and $\tau_{\omega}$ perpendicular to the detector plane.
$\tau_{\omega}$ is determined experimentally by recording the rocking
 curve, and corresponds to the rocking curve width.
}\label{fig:fig2new}
  \end{center}
\end{figure}

As shown in Fig. 1, the  appearance of magnetism is,
in essence, concomitant with the suppression of SANS intensity with field and underdoping.
Drawing upon results from the literature\citep{Gil02,Gil04,Cha06}
 and new observations reported herein, we are also able to plot the VL structure at low temperature $T$ versus magnetic field $H$
and doping $x$.

\section{Experimental Methods}

Single crystals of La$_{2-x}$Sr$_{x}$CuO$_{4}$ with $x$ = 0.105-0.22 were grown
 by the traveling solvent floating zone method.\citep{Nak98} The static and dynamic magnetic
properties of the samples were characterized using both neutron diffraction and neutron
spectroscopy and good agreement was found between
our data~\citep{Cha08,Cha09,Lip07}
and previously published results.\citep{Kha02,Kha03,Lak02,Kha05,Kat00}

The SANS experiments reported here were carried out over a series of experiments
using the SANS-I instrument at SINQ,\citep{Koh00} and the D11 and D22 instruments at ILL.
In all experiments we adopted the experimental geometry where external magnetic
fields $\mu_{0}H$ are applied parallel to the crystal \textbf{c}-axis, and almost parallel to
the neutron beam. The scattered neutrons are recorded using a position-sensitive-detector
placed behind the sample. For each doping, up to three different neutron wavelengths
spanning the range $\lambda_{n}$~=~5--16~\AA~were used in order to cover the applied
field $\mu_0H$ range of 0.03--10~T. In all cases, a zero-field cooled background was
 subtracted from the field-cooled data in order to leave just the VL signal.

Our experimental setup is shown in Fig.~\ref{fig:fig2new} where we also illustrate the relationship
between the VL in the sample and the quantities extracted at the position-sensitive-detector.
Typically, to observe the signal due to the VL, the sample and cryomagnet are rotated together
by angles (such as that shown by $\omega$ in Fig.~2) in order to bring a reciprocal lattice vector
onto the Bragg condition at the detector. Due to both the finite resolution of the instrument,
and the mosaic spread or imperfection of the VL, the Bragg spots occupy a finite volume in
reciprocal space, and can be described, in a first approximation, by three widths $w_{\ell}$, $w_{\rm q}$ and
$w_\perp$ as shown in Fig.~2. These three lengths correspond to the angular widths $\tau_\omega$, $\tau_{\rm r}$ and $\tau_{\rm A}$ respectively
measured in the SANS experiment.
$\tau_{\rm r}$ and $\tau_{\rm A}$ describe the finite size of the Bragg spot on the detector plane,
and $\tau_\omega$ is the width of a rocking curve,
measured by recording the Bragg spot intensity as a function of rotation angle $\omega$.
In the geometry shown in Fig.~2, $\tau_{\rm A}$ (and $w_\perp$) provide a measure of the VL orientational order about the field axis,
while $\tau_{\rm r}$ (and $w_{\rm q}$) is dominated by the SANS instrumental resolution function.
The contribution due to the resolution function is smallest for $\tau_\omega$ (and $w_{\ell}$),
which is sensitive to the VL correlation length along the field direction (vortex `straightness').
In each of our measuremenets,
the sample and the field were rotated together through a wide range of rocking angles $\omega$ spanning
$\pm$~5$^{\circ}$ about the neutron beam. Note that this angular rotation range is typically
narrower than the width of rocking curve in our samples. However, by summing over the
measured rocking angles, diffraction patterns such as those shown in
Fig. 3(a) and (b) are obtained, allowing a determination of the VL properties.

\begin{figure}
 \begin{center}
\includegraphics[width=0.46\textwidth]{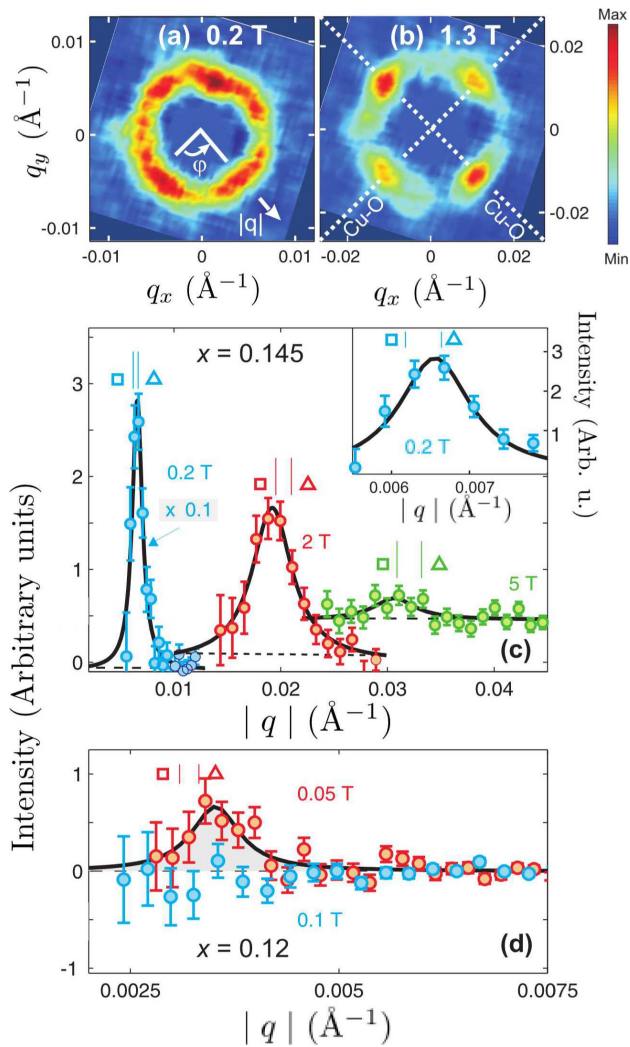}
\caption{(Color online) (a)-(b) Vortex lattice diffraction patterns, recorded
using a position sensitive detector (see Fig. 2), of LSCO $x=0.145$ under
applied magnetic fields of 0.2\,T and 1.3\,T, respectively.
Notice that the Cu-O axes are along the diagonal.
(c)-(d) Azimuthally-averaged momentum $|q|$-dependence
of the scattered intensity, summed over rocking angles,
for dopings $x=0.145$ and 0.12, and applied
magnetic fields as indicated.
For visibility, the 0.2~T data in (c) have been divided by a factor of ten.
Inset of (c) is a zoom on the 0.2~T data. The vertical bars above the
diffraction peaks indicate the expected positions for square and regular hexagonal VL coordinations.
The solid black lines in (c)-(d) are Lorentzian fits to the
data.
}\label{fig:fig1}
  \end{center}
\end{figure}

\section{Results and Discussion}
\subsection{Vortex lattice morphology}
Our observations of the VL structure and coordination
can be quantified in terms of a dimensionless parameter
$\sigma=4\pi^2 \mu_0H /(\Phi_0 |\bf{G}|^2 )$
where 
$|\bf{G}|$ is the magnitude of the reciprocal VL vector.
For a regular hexagonal VL coordination $\sigma=\sqrt{3}/2$,
while for a square coordination $\sigma=1$.
The definition of $\sigma$
is useful because it does
not require details of the positions $\bf{G}$ of Bragg peaks;
only the magnitude $|\bf{G}|$ is needed. At low fields, the VL is susceptible to
orientational disorder due to  impurities or defects in the sample,
but $|\bf{G}|$ can still be measured.
For example, in Fig.~\ref{fig:fig1}(a) the diffracted intensity measured in LSCO $x=0.145$ at
$0.2$~T does not show well-defined Bragg spots. This indicates a large $\tau_{\rm A}$ and poor
orientational order of the VL about the $c$-axis.  Nevertheless we can
determine the VL coordination by averaging over the azimuthal angle
$\varphi$ so that the diffracted intensity $I(|\bf{q}|)$ becomes a function of
$|\bf{q}|$ only,
see Fig.~\ref{fig:fig1}(c). $|\bf{G}|$ is determined from the peak position
in the $|\bf{q}|$-dependence. Fitting a Lorentzian  lineshape to these 0.2~T data,
yields $\sigma=0.88(2)$
very close to the value expected for a
hexagonal VL.
\begin{figure}
 \begin{center}
  \includegraphics[width=0.46\textwidth]{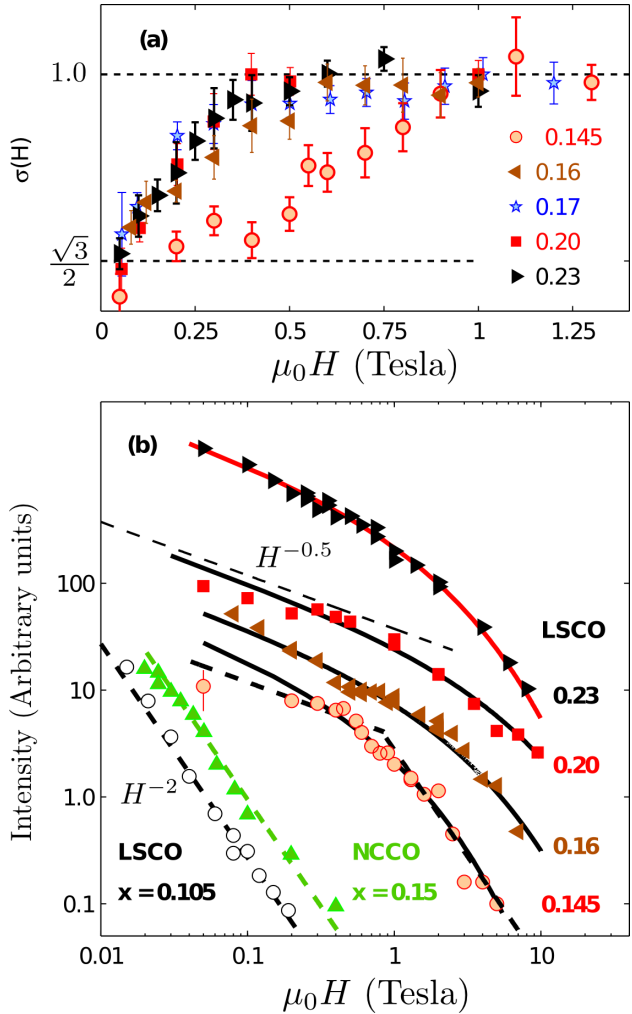}
  \caption{(Color online) (a) Dimensionless constant $\sigma$, defined in the text, as a function of
magnetic field for LSCO $x=0.145$, 0.16 $0.17$ (Ref.~\onlinecite{Gil02}),
0.20 (Ref.~\onlinecite{Gil04}), and 0.23. $\sigma=\sqrt{3}/2$ is expected for
a hexagonal lattice and a square vortex lattice has $\sigma=1$. The change form $\sigma=\sqrt{3}/2$
to 1 therefore reveals the a hexagonal-to-square transition of the vortex lattice structure (see also Fig.~1).
(b) SANS intensity $I$ versus applied magnetic field for NCCO $x=0.15$ (Ref.~\onlinecite{Gil04b})
and La$_{2-x}$Sr$_x$CuO$_4$ with $x=0.105$ (Ref.~\onlinecite{Cha08a}), and
 $0.145-0.23$,  (this work). For clarity, the intensities for each of the  compositions have been
given an arbitrary vertical
offset. Solid lines are fits to the Clem model form factor where the superconducting coherence
length $\xi$ is the only parameter (see Fig. 5). Dashed lines indicate power law dependencies; $I\sim H^{-0.5}$ and
$I\sim H^{-2}$. Notice that the Bragg glass paradigm for vortices in the presence of disorder  is consistent with a
cross over from  $I\sim H^{-0.5}$ to  $I\sim H^{-2}$, see text. }
 \label{fig:fig2}
 \end{center}
\end{figure}
At $\mu_0H=1.3$~T in Fig.~\ref{fig:fig1}(b), we find four Bragg spots with $\mathbf{{G}}$
along the Cu-O bond directions and $\sigma = 0.99(1)$,
indicating not only an improved VL orientational order but moreover a square coordination.
As shown quantitatively in Fig. 4(a), the VL coordination in underdoped LSCO $x=0.145$
changes steadily from hexagonal to square over the range $\mu_0H=0.2$ to 0.8~T.
 In contrast, on the optimally- and over-doped side of the phase diagram ($x\geq0.17$),
 the  VL structure becomes square by
$\mu_0 H\approx0.4$~T (Fig.~\ref{fig:fig2}(a)).

A minimum in either the Fermi velocity $v_F(\mathbf{k})$ or the superconducting gap $\Delta(\mathbf{k})$ are well-known
sources of field-driven hexagonal-to-square VL transitions.\citep{Nak02,Kei94,Bro04,Esk03}
 In both LSCO and YBa$_2$Cu$_3$O$_{y}$ (YBCO) the band structure is predominantly two-dimensional
 but  with some $c$-axis dispersion
 near the $\mathbf{k}= (\pi,0)$-point.\citep{Cha93,Yos06}
It was previously suggested that the VL morphology should be
understood from the Fermi surface topology near the ($\pi$,0)-point.\citep{Nak02}
 A crucial difference between the band structures of LSCO and YBCO  is
that in LSCO, a van Hove singularity crosses the Fermi level
($\epsilon_F$) at ($\pi$,0) somewhere between $x=0.17$ and
$0.22$,\citep{Yos06,Cha08b} leading to a huge $v_F$-anisotropy.
This offers an explanation as to why the square VL
is oriented along the ($\pi$,0) direction in LSCO as opposed to the
nodal ($\pi$,$\pi$)-direction in YBCO.\citep{Kei94,Bro04,Whi08,Whi09,Whi11,Nak02}
As LSCO is underdoped, the
 van Hove
singularity is pushed further away from $\epsilon_F$,\citep{Cha08b}
leading to a smaller
$v_F$-anisotropy and consequently
a larger field is required
to form the square VL, as indeed observed for LSCO $x=0.145$ (Fig. 4a).

\begin{figure}
 \begin{center}
  \includegraphics[width=0.46\textwidth]{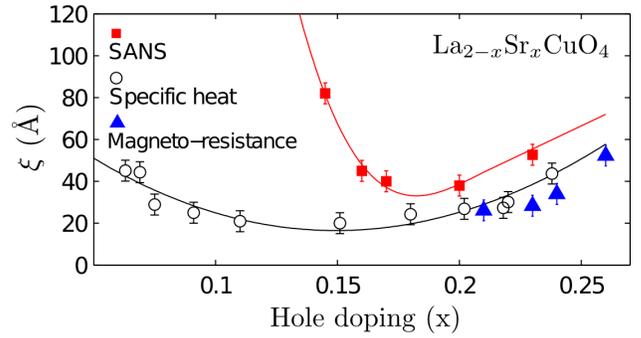}
  \caption{(Color online)  Superconducting coherence length $\xi$
in LSCO extracted from magneto-resistance (solid blue points),\citep{Rou11}
specific heat (open black points),\citep{Wan08}
and our SANS measurements (solid red points) and plotted
as a function of the hole concentration $x$. The high-field magneto-resistance study
measures the upper critical field $H_{c2}$ at low temperatures and we used
$H_{c2}=\Phi_0/(2\pi\xi^2)$ to estimate the coherence length $\xi$.
On the other hand, the specific heat and the SANS experiments were
carried out at fields smaller or comparable to $H_{c2}$. To extract
the superconducting coherence length from the SANS data
we used the Clem model for the VL form factor. Notice that
in presence of vortex lattice disorder, the Clem model  will overestimate the coherence
length (see text). This may explain why the coherence length extracted
from the SANS data lies systematically above the specific heat and
magneto-resistance measurements.
All lines are guides to the eye.
  }\label{fig:fig5}
 \end{center}
\end{figure}

\subsection{Diffracted SANS intensity}
We define the VL intensity of the first order diffraction peak $I$ as the sum of the area under $I(|\bf{q}|)$,
which is itself a sum over rocking angles.
Overlap measurements of  $I$ vs $\mu_0H$, shown in Fig. 4(b) for $x=0.105$--0.23,
were done whenever the neutron wavelength $\lambda_n$ was changed.
For $x\geq0.16$, intensity could be observed up to the highest
applied field $10$~T.
By contrast, for  $x = 0.145$
no intensity was observed above the quantum critical field
($ \mu_0 H=7\pm1$~T in our sample)
for SDW order.\citep{Kha05,Cha08,Cha09}
It is, however, still possible that the VL
extends slightly into the SDW ordered phase. This is the case
in LSCO $x=0.105$,\citep{Div04} where a 3D vortex lattice
exhibits $I\propto H^{-2}$ (Fig.~\ref{fig:fig2}(b)) over two decades of intensity and
co-exists with short range SDW order\citep{Cha08}
at very low fields $H_{c1}<\mu_0H \lesssim 0.2$~T.
VL intensity is also observed in LSCO $x=0.12$; a compound where
long-range SDW order
exists already in zero field.\citep{Cha08} At $\mu_0 H$~=~0.05~T, the
 $|\bf{q}|$-dependence of the  intensity  $I(|\bf{q}|)$ (Fig.~\ref{fig:fig1}(d)) suggests a
 hexagonal VL coordination.
 To the best of our knowledge, this
provides the first evidence by SANS
of a VL co-existing with SDW-order at the 1/8-anomaly.
The field range of co-existence is small; on increasing the
field to just $\mu_0 H$~=~0.1~T, Fig.~\ref{fig:fig1}(d) shows that the
VL signal has already fallen to the background level.
Notice that the fields 0.1--0.2 T are much smaller than
 those required to decouple 3D superconductivity.\citep{Sch10}

\subsection{Coherence length from the VL form factor}
In the case of perfect crystalline VL order,
the observed intensity of the first order diffraction peak
$I(H) \propto \sum \mathcal{F}^2/ |\bf{G}|$ where $\mathcal{F}$ is the form
factor of a single vortex and has units of field. The sum is over all the $\bf{q}$-vectors contributing
to the intensity near wave-vector $|\bf{G}|$, and we have assumed that the rocking curve width remains constant with field in obtaining $I(H)$.
A variational solution to the Ginzburg-Landau
model, namely the Clem model for the VL form factor,\citep{Cle75} yields $\mathcal{F}\propto GK_1(G\xi)$, where
$K_1$ denotes the modified Bessel function of first order, $G=|\mathbf{G}| = 2 \pi \sqrt{H/\sigma\Phi_0}$,
and  the vortex core size $\xi$ is the only fit parameter.\citep{Not01}
We point out that the application of the Clem model to our data
yields an upper bound for  $\xi$; disorder effects are, for example, not included.\citep{Not02}
With increasing vortex lattice disorder, the degree by which
the Clem model overestimates the coherence length is larger.
By comparing the doping dependence of the extracted $\xi$ 
with the
Ginzburg Landau
coherence length $\xi_{GL}=\sqrt{\Phi_0/2\pi H_{c2}}$ estimated indirectly
from specific heat\citep{Wan08} and high-field magneto-resistance
experiments,\citep{Rou11} a reasonable
 agreement is found on the overdoped side, see  Fig. 5.
This suggests that our SANS data indeed provide
a
measure of $\xi$, even though disorder is undoubtedly present.
 Identifying $\xi$ with the Pippard coherence
length   $\xi_p\sim\hbar v_F(k)/ \Delta(k)$ suggests
that the relatively  short coherence length
$\xi$ around optimally doped LSCO is not only due to
the large pairing gap $\Delta$; the small
Fermi velocity
is also playing a significant role.

\subsection{Disorder effects and structure factor}
On the underdoped side, we find a strong discrepancy
between the coherence length estimated from specific heat
and the SANS data fitted with the Clem model for the form factor, see Fig. 5. For  LSCO $x=0.145$,  we find
$\xi\sim 80$~\AA\ corresponding to an unrealistically small
upper critical field $\mu_0 H_{c2}=\Phi_0/ 2\pi \xi^2\sim5$~T.
A larger coherence length may result from the weakening of superconductivity
due to competition with,
for example, magnetism.
However, this does not explain
the discrepancy between our SANS data and the specific
heat data.\citep{Wan08}
A more plausible explanation  is
that the VL disorder potential increases with underdoping.
It is possible  that VL disorder proliferates as the system
is tuned towards the state where magnetism and superconductivity coexist.

The VL displacements throughout the doping range are well-described
by elastic theory, namely the Bragg glass (BrG) paradigm.\citep{Gia95,Bog01}
In the presence of disorder, the positional order of an elastic VL
decays exponentially with a characteristic length scale $R_A$.
(With increasing disorder  $R_A\rightarrow0$.)
If such an exponential decay were to persist at all
length scales $R$,
a total destruction of long-range order would result.\citep{Lar70}
This proposed destruction, even under weak disorder, presented a long-standing
puzzle with respect to experimental observations
where Bragg peaks may readily be observed.
Theoretically, the puzzle was resolved with the advent of the BrG paradigm, in
which an asymptotic regime for $R>R_A$ enters,
where the positional order decays only weakly,
 leading to algebraically diverging Bragg peaks.
At the crossover scale $R_A$ vortex displacements are comparable to the lattice
spacing $a_0=\sqrt{\sigma\Phi_0/H}$.
In earlier muon spin rotation ($\mu$SR) work on a LSCO $x=0.105$ sample,
an order-disorder transition was observed and
associated with a transition out of the quasi-long-range-ordered BrG phase.\citep{Div04}
VL correlations in the BrG phase have been explored more directly in
a study of low-purity niobium.\citep{Lav08}
We point out that the BrG paradigm can explain
both the $H^{-0.5}$ and $H^{-2}$ field-dependences of intensity (c.f. Fig. 4(b)),
as well as the crossover between them.
Dependent on whether
 the instrumental resolution $s\approx 60 a_0$ is larger or smaller than
 $R_A$, a different field dependence
 of the SANS intensity is predicted.\citep{Kle01}
 For $s<R_A$, the contribution to the structure factor is
 identical to that of a crystalline VL, hence $I\sim 1/\sqrt{H}$.
 In the other limit $s>R_A$, an additional factor $H^{-\mu}$
 contributes to the 
 intensity.  Elastic theory\citep{Gia95} yields
 $\mu=3/2$ and hence $I\approx H^{-2}$ --- as
 indeed we observed previously in LSCO $x=0.105$,\citep{Cha08a} see Fig.~4b.
The intensity for
 $x=0.145$ is also consistent with a
 $H^{-0.5}$ to $H^{-2}$ crossover, see Fig.~3.
 At the crossover field ($\sim 0.85$~T),
 $R_A\approx s \approx 60 a_0 \approx 3~\mu$m.
 By contrast, in LSCO $x=0.105$
 the crossover field $\ll 0.05$~T,
implying that the disorder potential
 increases dramatically with decreasing doping.
This disordering seems to occur as the static magnetic
correlation length $\zeta$ approaches the VL spacing ($\zeta\rightarrow a_0$),
suggesting an electronic origin to the VL disorder effectuated
by real space competition between $\zeta$ and $a_0$, rather than orthorhombic
twin boundaries or impurities as observed in other superconductors.\citep{Lav08}
We noticed  that the BrG model also provides an excellent
description for the  (previously unexplained)  field dependence of intensity
in Nd$_{2-x}$Ce$_x$CuO$_4$ (NCCO) with
 $x=0.15$ (Fig.~4(b)).\citep{Gil04b} More experimental studies are required on NCCO
to establish if the VL disorder therein has origins similar to that
in LSCO.

\subsection{Real space picture}
We now consider how magnetic and superconducting order parameters might
coexist in real space. Around $x=0.12$ doping in LSCO,  $\mu$SR
measurements revealed magnetic and superconducting fractions that together exceed 100$\%$.
It was therefore concluded that the magnetic and superconducting order parameters
are not phase separated but rather intertwined on a nanometer scale.\citep{Sav02,Uem03}
This real space picture is not easily reconciled with neutron diffraction
studies that report a magnetic correlation length of several hundreds
of Angstroms.\citep{Cha08,Cha09,Kha02,Kha03}
One possibility is that the charge of the
muon induces magnetism in which
case the $\mu$SR technique  overestimates the magnetic volume
fraction.\citep{Dan10}
Here we showed that the vortex lattice becomes more
disordered as the vortex interspacing approaches the magnetic
correlation length. This suggests that the magnetic and superconducting
order parameters are coupled. How magnetism, superconductivity, and
 vortices are arranged in real space when the spin correlation length
is larger than the vortex spacing ($\zeta>a_0$) is an interesting
question that is difficult to address with the SANS technique since
no observable SANS signal is found in that region of the phase diagram,
see Fig. 1.  Direct imaging techniques\citep{Vin05,Blu06,Mol11} are undoubtedly more informative
in this regime although it may be experimentally challenging to probe magnetism and
vortices simultaneously.

\section{Conclusions}

In summary, our studies of the vortex lattice in LSCO allow us
to draw two main conclusions. First, near optimal doping, and far
from the SDW instability, the VL structure/orientation and the
small superconducting coherence length $\xi$ (and hence large upper
critical field $H_{c2}$) may both be rationalized as arising from
a vanishing Fermi velocity
due to the van-Hove singularity near the zone boundary.
 Second, we find that the fermiological picture breaks down as the SDW
instability is approached by tuning either the doping or the applied magnetic field.
There, the vortex lattice structure factor needs to be
accounted for in the field-dependence of the observable
SANS intensity.
The Bragg glass paradigm, describing vortex lattices in the
presence of weak disorder, accounts for the SANS intensity
behavior across the entire phase diagram (where a SANS signal
is discernible), from the underdoped $x=$0.105, to the optimally
doped $x~\sim$0.16 and overdoped $x=$0.23 regimes. In particular, it
is able to explain the cross-over as the SDW instability is
approached. Evidently, the SDW order acts as a novel
electronic provenience of disorder on the vortex lattice in LSCO.

Experiments were performed at the Institut Laue-Langevin (ILL),
Grenoble, France and at the Swiss Spallation Source SINQ, Paul
Scherrer Institute, Villigen, Switzerland. We acknowledge discussions with C. Niedermayer,
B.M. Andersen, and N.B. Christensen and support
from the Swiss NSF (through NCCR, MaNEP, and grant Nr 200020-105151,
PBEZP2-122855), DanScatt, UK EPSRC, and by the Ministry of Education
and Science of Japan.

\bibliography{LSCO}

\begin{thebibliography}{60}%
\makeatletter
\providecommand \@ifxundefined [1]{%
 \@ifx{#1\undefined}
}%
\providecommand \@ifnum [1]{%
 \ifnum #1\expandafter \@firstoftwo
 \else \expandafter \@secondoftwo
 \fi
}%
\providecommand \@ifx [1]{%
 \ifx #1\expandafter \@firstoftwo
 \else \expandafter \@secondoftwo
 \fi
}%
\providecommand \natexlab [1]{#1}%
\providecommand \enquote  [1]{``#1''}%
\providecommand \bibnamefont  [1]{#1}%
\providecommand \bibfnamefont [1]{#1}%
\providecommand \citenamefont [1]{#1}%
\providecommand \href@noop [0]{\@secondoftwo}%
\providecommand \href [0]{\begingroup \@sanitize@url \@href}%
\providecommand \@href[1]{\@@startlink{#1}\@@href}%
\providecommand \@@href[1]{\endgroup#1\@@endlink}%
\providecommand \@sanitize@url [0]{\catcode `\\12\catcode `\$12\catcode
  `\&12\catcode `\#12\catcode `\^12\catcode `\_12\catcode `\%12\relax}%
\providecommand \@@startlink[1]{}%
\providecommand \@@endlink[0]{}%
\providecommand \url  [0]{\begingroup\@sanitize@url \@url }%
\providecommand \@url [1]{\endgroup\@href {#1}{\urlprefix }}%
\providecommand \urlprefix  [0]{URL }%
\providecommand \Eprint [0]{\href }%
\@ifxundefined \urlstyle {%
  \providecommand \doi  [0]{\begingroup \@sanitize@url \@doi}%
  \providecommand \@doi [1]{\endgroup \@@startlink {\doibase
  #1}doi:\discretionary {}{}{}#1\@@endlink }%
}{%
  \providecommand \doi  [0]{doi:\discretionary{}{}{}\begingroup
  \urlstyle{rm}\Url }%
}%
\providecommand \doibase [0]{http://dx.doi.org/}%
\providecommand \Doi [0]{\begingroup \@sanitize@url \@Doi }%
\providecommand \@Doi  [1]{\endgroup\@@startlink{\doibase#1}\@@Doi}%
\providecommand \@@Doi [1]{#1\@@endlink}%
\providecommand \selectlanguage [0]{\@gobble}%
\providecommand \bibinfo  [0]{\@secondoftwo}%
\providecommand \bibfield  [0]{\@secondoftwo}%
\providecommand \translation [1]{[#1]}%
\providecommand \BibitemOpen [0]{}%
\providecommand \bibitemStop [0]{}%
\providecommand \bibitemNoStop [0]{.\EOS\space}%
\providecommand \EOS [0]{\spacefactor3000\relax}%
\providecommand \BibitemShut  [1]{\csname bibitem#1\endcsname}%
\bibitem [{\citenamefont {Canfield}\ \emph {et~al.}(1998)\citenamefont
  {Canfield}, \citenamefont {Gammel},\ and\ \citenamefont {Bisop}}]{Can98}%
  \BibitemOpen
  \bibfield  {author} {\bibinfo {author} {\bibfnamefont {P.~C.}\ \bibnamefont
  {Canfield}}, \bibinfo {author} {\bibfnamefont {P.~L.}\ \bibnamefont
  {Gammel}}, \ and\ \bibinfo {author} {\bibfnamefont {D.~J.}\ \bibnamefont
  {Bisop}},\ }\href@noop {} {\bibfield  {journal} {\bibinfo  {journal} {Phys.
  Today},\ }\textbf {\bibinfo {volume} {51}},\ \bibinfo {pages} {40} (\bibinfo
  {year} {1998})}\BibitemShut {NoStop}%
\bibitem [{\citenamefont {M\"{u}ller}\ and\ \citenamefont
  {Narozhnyi}(2001)}]{Mul01}%
  \BibitemOpen
  \bibfield  {author} {\bibinfo {author} {\bibfnamefont {K.-H.}\ \bibnamefont
  {M\"{u}ller}}\ and\ \bibinfo {author} {\bibfnamefont {V.~N.}\ \bibnamefont
  {Narozhnyi}},\ }\href@noop {} {\bibfield  {journal} {\bibinfo  {journal}
  {Rep. Prog. Phys.},\ }\textbf {\bibinfo {volume} {64}},\ \bibinfo {pages}
  {943} (\bibinfo {year} {2001})}\BibitemShut {NoStop}%
\bibitem [{\citenamefont {Gupta}(2006)}]{Gup06}%
  \BibitemOpen
  \bibfield  {author} {\bibinfo {author} {\bibfnamefont {L.~C.}\ \bibnamefont
  {Gupta}},\ }\href@noop {} {\bibfield  {journal} {\bibinfo  {journal}
  {Advances in Physics},\ }\textbf {\bibinfo {volume} {55}},\ \bibinfo {pages}
  {691} (\bibinfo {year} {2006})}\BibitemShut {NoStop}%
\bibitem [{\citenamefont {Budko}\ and\ \citenamefont {Canfield}(2006)}]{Bud06}%
  \BibitemOpen
  \bibfield  {author} {\bibinfo {author} {\bibfnamefont {S.~L.}\ \bibnamefont
  {Budko}}\ and\ \bibinfo {author} {\bibfnamefont {P.~C.}\ \bibnamefont
  {Canfield}},\ }\href@noop {} {\bibfield  {journal} {\bibinfo  {journal} {C.
  R. Physique},\ }\textbf {\bibinfo {volume} {7}},\ \bibinfo {pages} {56}
  (\bibinfo {year} {2006})}\BibitemShut {NoStop}%
\bibitem [{\citenamefont {Lake}\ \emph {et~al.}(2002)\citenamefont {Lake},
  \citenamefont {R{\o}nnow}, \citenamefont {Christensen}, \citenamefont
  {Aeppli}, \citenamefont {Lefmann}, \citenamefont {McMorrow}, \citenamefont
  {Vorderwisch}, \citenamefont {Smeibidl}, \citenamefont {Mangkorntong},
  \citenamefont {Sasagawa}, \citenamefont {Nohara}, \citenamefont {Takagi},\
  and\ \citenamefont {Mason}}]{Lak02}%
  \BibitemOpen
  \bibfield  {author} {\bibinfo {author} {\bibfnamefont {B.}~\bibnamefont
  {Lake}}, \bibinfo {author} {\bibfnamefont {H.~M.}\ \bibnamefont {R{\o}nnow}},
  \bibinfo {author} {\bibfnamefont {N.~B.}\ \bibnamefont {Christensen}},
  \bibinfo {author} {\bibfnamefont {G.}~\bibnamefont {Aeppli}}, \bibinfo
  {author} {\bibfnamefont {K.}~\bibnamefont {Lefmann}}, \bibinfo {author}
  {\bibfnamefont {D.~F.}\ \bibnamefont {McMorrow}}, \bibinfo {author}
  {\bibfnamefont {P.}~\bibnamefont {Vorderwisch}}, \bibinfo {author}
  {\bibfnamefont {P.}~\bibnamefont {Smeibidl}}, \bibinfo {author}
  {\bibfnamefont {N.}~\bibnamefont {Mangkorntong}}, \bibinfo {author}
  {\bibfnamefont {T.}~\bibnamefont {Sasagawa}}, \bibinfo {author}
  {\bibfnamefont {M.}~\bibnamefont {Nohara}}, \bibinfo {author} {\bibfnamefont
  {H.}~\bibnamefont {Takagi}}, \ and\ \bibinfo {author} {\bibfnamefont {T.~E.}\
  \bibnamefont {Mason}},\ }\href@noop {} {\bibfield  {journal} {\bibinfo
  {journal} {Nature},\ }\textbf {\bibinfo {volume} {415}},\ \bibinfo {pages}
  {299} (\bibinfo {year} {2002})}\BibitemShut {NoStop}%
\bibitem [{\citenamefont {Eskildsen}\ \emph {et~al.}(2011)\citenamefont
  {Eskildsen}, \citenamefont {Forgan},\ and\ \citenamefont
  {Kawano-Furukawa}}]{Esk11}%
  \BibitemOpen
  \bibfield  {author} {\bibinfo {author} {\bibfnamefont {M.~R.}\ \bibnamefont
  {Eskildsen}}, \bibinfo {author} {\bibfnamefont {E.~M.}\ \bibnamefont
  {Forgan}}, \ and\ \bibinfo {author} {\bibfnamefont {H.}~\bibnamefont
  {Kawano-Furukawa}},\ }\href@noop {} {\bibfield  {journal} {\bibinfo
  {journal} {Rep. Prog. Phys.},\ }\textbf {\bibinfo {volume} {74}},\ \bibinfo
  {pages} {124504} (\bibinfo {year} {2011})}\BibitemShut {NoStop}%
\bibitem [{\citenamefont {Monthoux}\ \emph {et~al.}(2007)\citenamefont
  {Monthoux}, \citenamefont {Pines},\ and\ \citenamefont {Lonzarich}}]{Mon07}%
  \BibitemOpen
  \bibfield  {author} {\bibinfo {author} {\bibfnamefont {P.}~\bibnamefont
  {Monthoux}}, \bibinfo {author} {\bibfnamefont {D.}~\bibnamefont {Pines}}, \
  and\ \bibinfo {author} {\bibfnamefont {G.~G.}\ \bibnamefont {Lonzarich}},\
  }\href@noop {} {\bibfield  {journal} {\bibinfo  {journal} {Nature},\ }\textbf
  {\bibinfo {volume} {450}},\ \bibinfo {pages} {1177} (\bibinfo {year}
  {2007})}\BibitemShut {NoStop}%
\bibitem [{\citenamefont {J\'{e}rome}(1983)}]{Jer83}%
  \BibitemOpen
  \bibfield  {author} {\bibinfo {author} {\bibfnamefont {D.}~\bibnamefont
  {J\'{e}rome}},\ }\href@noop {} {\bibfield  {journal} {\bibinfo  {journal}
  {Journal of Magnetism and Magnetic Materials},\ }\textbf {\bibinfo {volume}
  {31–34}},\ \bibinfo {pages} {20–28} (\bibinfo {year} {1983})}\BibitemShut
  {NoStop}%
\bibitem [{\citenamefont {Vojta}(2009)}]{Voj09}%
  \BibitemOpen
  \bibfield  {author} {\bibinfo {author} {\bibfnamefont {M.}~\bibnamefont
  {Vojta}},\ }\href@noop {} {\bibfield  {journal} {\bibinfo  {journal} {Adv.
  Phys.},\ }\textbf {\bibinfo {volume} {58}},\ \bibinfo {pages} {699} (\bibinfo
  {year} {2009})}\BibitemShut {NoStop}%
\bibitem [{\citenamefont {Demler}\ \emph {et~al.}(2001)\citenamefont {Demler},
  \citenamefont {Sachdev},\ and\ \citenamefont {Zhang}}]{Dem01}%
  \BibitemOpen
  \bibfield  {author} {\bibinfo {author} {\bibfnamefont {E.}~\bibnamefont
  {Demler}}, \bibinfo {author} {\bibfnamefont {S.}~\bibnamefont {Sachdev}}, \
  and\ \bibinfo {author} {\bibfnamefont {Y.}~\bibnamefont {Zhang}},\
  }\href@noop {} {\bibfield  {journal} {\bibinfo  {journal} {Phys. Rev.
  Lett.},\ }\textbf {\bibinfo {volume} {87}},\ \bibinfo {pages} {067202}
  (\bibinfo {year} {2001})}\BibitemShut {NoStop}%
\bibitem [{\citenamefont {Hoffmann}\ \emph {et~al.}(2002)\citenamefont
  {Hoffmann}, \citenamefont {Hudson}, \citenamefont {Lang}, \citenamefont
  {Madhavan}, \citenamefont {Eisaki}, \citenamefont {Uchida},\ and\
  \citenamefont {Davis}}]{Hof02}%
  \BibitemOpen
  \bibfield  {author} {\bibinfo {author} {\bibfnamefont {J.}~\bibnamefont
  {Hoffmann}}, \bibinfo {author} {\bibfnamefont {E.~W.}\ \bibnamefont
  {Hudson}}, \bibinfo {author} {\bibfnamefont {K.~M.}\ \bibnamefont {Lang}},
  \bibinfo {author} {\bibfnamefont {V.}~\bibnamefont {Madhavan}}, \bibinfo
  {author} {\bibfnamefont {H.}~\bibnamefont {Eisaki}}, \bibinfo {author}
  {\bibfnamefont {S.}~\bibnamefont {Uchida}}, \ and\ \bibinfo {author}
  {\bibfnamefont {J.~C.}\ \bibnamefont {Davis}},\ }\href@noop {} {\bibfield
  {journal} {\bibinfo  {journal} {Science},\ }\textbf {\bibinfo {volume}
  {295}},\ \bibinfo {pages} {466} (\bibinfo {year} {2002})}\BibitemShut
  {NoStop}%
\bibitem [{\citenamefont {Sachdev}(2003)}]{Sac03}%
  \BibitemOpen
  \bibfield  {author} {\bibinfo {author} {\bibfnamefont {S.}~\bibnamefont
  {Sachdev}},\ }\href@noop {} {\bibfield  {journal} {\bibinfo  {journal} {Rev.
  Mod. Phys.},\ }\textbf {\bibinfo {volume} {75}},\ \bibinfo {pages} {913}
  (\bibinfo {year} {2003})}\BibitemShut {NoStop}%
\bibitem [{\citenamefont {Andersen}\ \emph {et~al.}(2009)\citenamefont
  {Andersen}, \citenamefont {Sylj{\aa}sen},\ and\ \citenamefont
  {Hedeg{\aa}rd}}]{And09}%
  \BibitemOpen
  \bibfield  {author} {\bibinfo {author} {\bibfnamefont {B.~M.}\ \bibnamefont
  {Andersen}}, \bibinfo {author} {\bibfnamefont {O.~F.}\ \bibnamefont
  {Sylj{\aa}sen}}, \ and\ \bibinfo {author} {\bibfnamefont {P.}~\bibnamefont
  {Hedeg{\aa}rd}},\ }\href@noop {} {\bibfield  {journal} {\bibinfo  {journal}
  {Phys. Rev. B},\ }\textbf {\bibinfo {volume} {80}},\ \bibinfo {pages}
  {052509} (\bibinfo {year} {2009})}\BibitemShut {NoStop}%
\bibitem [{\citenamefont {Lake}\ \emph {et~al.}(2001)\citenamefont {Lake},
  \citenamefont {Aeppli}, \citenamefont {Clausen}, \citenamefont {McMorrow},
  \citenamefont {Lefmann}, \citenamefont {Hussey}, \citenamefont
  {Mangkorntong}, \citenamefont {Nohara}, \citenamefont {Takagi}, \citenamefont
  {Mason},\ and\ \citenamefont {Schr\"{o}der}}]{Lak01}%
  \BibitemOpen
  \bibfield  {author} {\bibinfo {author} {\bibfnamefont {B.}~\bibnamefont
  {Lake}}, \bibinfo {author} {\bibfnamefont {G.}~\bibnamefont {Aeppli}},
  \bibinfo {author} {\bibfnamefont {K.~N.}\ \bibnamefont {Clausen}}, \bibinfo
  {author} {\bibfnamefont {D.~F.}\ \bibnamefont {McMorrow}}, \bibinfo {author}
  {\bibfnamefont {K.}~\bibnamefont {Lefmann}}, \bibinfo {author} {\bibfnamefont
  {N.~E.}\ \bibnamefont {Hussey}}, \bibinfo {author} {\bibfnamefont
  {N.}~\bibnamefont {Mangkorntong}}, \bibinfo {author} {\bibfnamefont
  {M.}~\bibnamefont {Nohara}}, \bibinfo {author} {\bibfnamefont
  {H.}~\bibnamefont {Takagi}}, \bibinfo {author} {\bibfnamefont {T.~E.}\
  \bibnamefont {Mason}}, \ and\ \bibinfo {author} {\bibfnamefont
  {A.}~\bibnamefont {Schr\"{o}der}},\ }\href@noop {} {\bibfield  {journal}
  {\bibinfo  {journal} {Science},\ }\textbf {\bibinfo {volume} {291}},\
  \bibinfo {pages} {1759} (\bibinfo {year} {2001})}\BibitemShut {NoStop}%
\bibitem [{\citenamefont {Shekhter}\ \emph {et~al.}(2011)\citenamefont
  {Shekhter}, \citenamefont {Bulaevskii},\ and\ \citenamefont
  {Batista}}]{She11}%
  \BibitemOpen
  \bibfield  {author} {\bibinfo {author} {\bibfnamefont {A.}~\bibnamefont
  {Shekhter}}, \bibinfo {author} {\bibfnamefont {L.~N.}\ \bibnamefont
  {Bulaevskii}}, \ and\ \bibinfo {author} {\bibfnamefont {C.~D.}\ \bibnamefont
  {Batista}},\ }\href@noop {} {\bibfield  {journal} {\bibinfo  {journal} {Phys.
  Rev. Lett.},\ }\textbf {\bibinfo {volume} {106}},\ \bibinfo {pages} {037001}
  (\bibinfo {year} {2011})}\BibitemShut {NoStop}%
\bibitem [{\citenamefont {Giamarchi}\ and\ \citenamefont
  {Le~Doussal}(1995)}]{Gia95}%
  \BibitemOpen
  \bibfield  {author} {\bibinfo {author} {\bibfnamefont {T.}~\bibnamefont
  {Giamarchi}}\ and\ \bibinfo {author} {\bibfnamefont {P.}~\bibnamefont
  {Le~Doussal}},\ }\href@noop {} {\bibfield  {journal} {\bibinfo  {journal}
  {Phys. Rev. B},\ }\textbf {\bibinfo {volume} {52}},\ \bibinfo {pages} {1242}
  (\bibinfo {year} {1995})}\BibitemShut {NoStop}%
\bibitem [{\citenamefont {Bogner}\ and\ \citenamefont
  {Nattermann}(2001)}]{Bog01}%
  \BibitemOpen
  \bibfield  {author} {\bibinfo {author} {\bibfnamefont {T.}~\bibnamefont
  {Bogner}, \bibfnamefont {S.~amd~Emig}}\ and\ \bibinfo {author} {\bibfnamefont
  {T.}~\bibnamefont {Nattermann}},\ }\href@noop {} {\bibfield  {journal}
  {\bibinfo  {journal} {Phys. Rev. B},\ }\textbf {\bibinfo {volume} {63}},\
  \bibinfo {pages} {174501} (\bibinfo {year} {2001})}\BibitemShut {NoStop}%
\bibitem [{\citenamefont {Vinnikov}\ \emph {et~al.}(2005)\citenamefont
  {Vinnikov}, \citenamefont {Anderegg}, \citenamefont {Bud'ko}, \citenamefont
  {Canfield},\ and\ \citenamefont {Kogan}}]{Vin05}%
  \BibitemOpen
  \bibfield  {author} {\bibinfo {author} {\bibfnamefont {L.~Y.}\ \bibnamefont
  {Vinnikov}}, \bibinfo {author} {\bibfnamefont {J.}~\bibnamefont {Anderegg}},
  \bibinfo {author} {\bibfnamefont {S.~L.}\ \bibnamefont {Bud'ko}}, \bibinfo
  {author} {\bibfnamefont {P.~C.}\ \bibnamefont {Canfield}}, \ and\ \bibinfo
  {author} {\bibfnamefont {V.~G.}\ \bibnamefont {Kogan}},\ }\href@noop {}
  {\bibfield  {journal} {\bibinfo  {journal} {Phys. Rev. B},\ }\textbf
  {\bibinfo {volume} {71}},\ \bibinfo {pages} {224513} (\bibinfo {year}
  {2005})}\BibitemShut {NoStop}%
\bibitem [{\citenamefont {Bluhm}\ \emph {et~al.}(2006)\citenamefont {Bluhm},
  \citenamefont {Sebastian}, \citenamefont {Guikema}, \citenamefont {Fisher},\
  and\ \citenamefont {Moler}}]{Blu06}%
  \BibitemOpen
  \bibfield  {author} {\bibinfo {author} {\bibfnamefont {H.}~\bibnamefont
  {Bluhm}}, \bibinfo {author} {\bibfnamefont {S.~E.}\ \bibnamefont
  {Sebastian}}, \bibinfo {author} {\bibfnamefont {J.~W.}\ \bibnamefont
  {Guikema}}, \bibinfo {author} {\bibfnamefont {I.~R.}\ \bibnamefont {Fisher}},
  \ and\ \bibinfo {author} {\bibfnamefont {K.~A.}\ \bibnamefont {Moler}},\
  }\href@noop {} {\bibfield  {journal} {\bibinfo  {journal} {Phys. Rev. B},\
  }\textbf {\bibinfo {volume} {73}},\ \bibinfo {pages} {014514} (\bibinfo
  {year} {2006})}\BibitemShut {NoStop}%
\bibitem [{\citenamefont {Kalisky}\ \emph {et~al.}(2011)\citenamefont
  {Kalisky}, \citenamefont {Kirtley}, \citenamefont {Analytis}, \citenamefont
  {Chu}, \citenamefont {Fisher},\ and\ \citenamefont {Moler}}]{Mol11}%
  \BibitemOpen
  \bibfield  {author} {\bibinfo {author} {\bibfnamefont {B.}~\bibnamefont
  {Kalisky}}, \bibinfo {author} {\bibfnamefont {J.~R.}\ \bibnamefont
  {Kirtley}}, \bibinfo {author} {\bibfnamefont {J.~G.}\ \bibnamefont
  {Analytis}}, \bibinfo {author} {\bibfnamefont {J.-H.}\ \bibnamefont {Chu}},
  \bibinfo {author} {\bibfnamefont {I.~R.}\ \bibnamefont {Fisher}}, \ and\
  \bibinfo {author} {\bibfnamefont {K.~A.}\ \bibnamefont {Moler}},\ }\href@noop
  {} {\bibfield  {journal} {\bibinfo  {journal} {Phys. Rev. B},\ }\textbf
  {\bibinfo {volume} {83}},\ \bibinfo {pages} {064511} (\bibinfo {year}
  {2011})}\BibitemShut {NoStop}%
\bibitem [{\citenamefont {Laver}\ \emph {et~al.}(2008)\citenamefont {Laver},
  \citenamefont {Forgan}, \citenamefont {Abrahamsen}, \citenamefont {Bowell},
  \citenamefont {Geue},\ and\ \citenamefont {Cubitt}}]{Lav08}%
  \BibitemOpen
  \bibfield  {author} {\bibinfo {author} {\bibfnamefont {M.}~\bibnamefont
  {Laver}}, \bibinfo {author} {\bibfnamefont {E.~M.}\ \bibnamefont {Forgan}},
  \bibinfo {author} {\bibfnamefont {A.~B.}\ \bibnamefont {Abrahamsen}},
  \bibinfo {author} {\bibfnamefont {C.}~\bibnamefont {Bowell}}, \bibinfo
  {author} {\bibfnamefont {T.}~\bibnamefont {Geue}}, \ and\ \bibinfo {author}
  {\bibfnamefont {R.}~\bibnamefont {Cubitt}},\ }\href@noop {} {\bibfield
  {journal} {\bibinfo  {journal} {Phys. Rev. Lett.},\ }\textbf {\bibinfo
  {volume} {100}},\ \bibinfo {pages} {107001} (\bibinfo {year}
  {2008})}\BibitemShut {NoStop}%
\bibitem [{\citenamefont {Divakar}\ \emph {et~al.}(2004)\citenamefont
  {Divakar}, \citenamefont {Drew}, \citenamefont {Lee}, \citenamefont
  {Gilardi}, \citenamefont {Mesot}, \citenamefont {Ogrin}, \citenamefont
  {Charalambous}, \citenamefont {Forgan}, \citenamefont {Menon}, \citenamefont
  {Momono}, \citenamefont {Oda}, \citenamefont {Dewhurst},\ and\ \citenamefont
  {Baines}}]{Div04}%
  \BibitemOpen
  \bibfield  {author} {\bibinfo {author} {\bibfnamefont {U.}~\bibnamefont
  {Divakar}}, \bibinfo {author} {\bibfnamefont {A.~J.}\ \bibnamefont {Drew}},
  \bibinfo {author} {\bibfnamefont {S.~L.}\ \bibnamefont {Lee}}, \bibinfo
  {author} {\bibfnamefont {R.}~\bibnamefont {Gilardi}}, \bibinfo {author}
  {\bibfnamefont {J.}~\bibnamefont {Mesot}}, \bibinfo {author} {\bibfnamefont
  {F.~Y.}\ \bibnamefont {Ogrin}}, \bibinfo {author} {\bibfnamefont
  {D.}~\bibnamefont {Charalambous}}, \bibinfo {author} {\bibfnamefont {E.~M.}\
  \bibnamefont {Forgan}}, \bibinfo {author} {\bibfnamefont {G.~I.}\
  \bibnamefont {Menon}}, \bibinfo {author} {\bibfnamefont {N.}~\bibnamefont
  {Momono}}, \bibinfo {author} {\bibfnamefont {M.}~\bibnamefont {Oda}},
  \bibinfo {author} {\bibfnamefont {C.~D.}\ \bibnamefont {Dewhurst}}, \ and\
  \bibinfo {author} {\bibfnamefont {C.}~\bibnamefont {Baines}},\ }\href@noop {}
  {\bibfield  {journal} {\bibinfo  {journal} {Phys. Rev. Lett.},\ }\textbf
  {\bibinfo {volume} {92}},\ \bibinfo {pages} {237004} (\bibinfo {year}
  {2004})}\BibitemShut {NoStop}%
\bibitem [{\citenamefont {Gilardi}\ \emph {et~al.}(2002)\citenamefont
  {Gilardi}, \citenamefont {Mesot}, \citenamefont {Drew}, \citenamefont
  {Divakar}, \citenamefont {Lee}, \citenamefont {Forgan}, \citenamefont
  {Zaharko}, \citenamefont {Conder}, \citenamefont {Aswal}, \citenamefont
  {Dewhurst}, \citenamefont {Cubitt}, \citenamefont {Momono},\ and\
  \citenamefont {Oda}}]{Gil02}%
  \BibitemOpen
  \bibfield  {author} {\bibinfo {author} {\bibfnamefont {R.}~\bibnamefont
  {Gilardi}}, \bibinfo {author} {\bibfnamefont {J.}~\bibnamefont {Mesot}},
  \bibinfo {author} {\bibfnamefont {A.}~\bibnamefont {Drew}}, \bibinfo {author}
  {\bibfnamefont {U.}~\bibnamefont {Divakar}}, \bibinfo {author} {\bibfnamefont
  {S.~L.}\ \bibnamefont {Lee}}, \bibinfo {author} {\bibfnamefont {E.~M.}\
  \bibnamefont {Forgan}}, \bibinfo {author} {\bibfnamefont {O.}~\bibnamefont
  {Zaharko}}, \bibinfo {author} {\bibfnamefont {K.}~\bibnamefont {Conder}},
  \bibinfo {author} {\bibfnamefont {V.~K.}\ \bibnamefont {Aswal}}, \bibinfo
  {author} {\bibfnamefont {C.~D.}\ \bibnamefont {Dewhurst}}, \bibinfo {author}
  {\bibfnamefont {R.}~\bibnamefont {Cubitt}}, \bibinfo {author} {\bibfnamefont
  {N.}~\bibnamefont {Momono}}, \ and\ \bibinfo {author} {\bibfnamefont
  {M.}~\bibnamefont {Oda}},\ }\href@noop {} {\bibfield  {journal} {\bibinfo
  {journal} {Phys. Rev. Lett.},\ }\textbf {\bibinfo {volume} {88}},\ \bibinfo
  {pages} {217003} (\bibinfo {year} {2002})}\BibitemShut {NoStop}%
\bibitem [{\citenamefont {Gilardi}\ \emph
  {et~al.}(2004){\natexlab{a}}\citenamefont {Gilardi}, \citenamefont {Mesot},
  \citenamefont {Drew}, \citenamefont {Divakar}, \citenamefont {Lee},
  \citenamefont {Andersen}, \citenamefont {Kohlbrecher}, \citenamefont
  {Momono},\ and\ \citenamefont {Oda}}]{Gil04}%
  \BibitemOpen
  \bibfield  {author} {\bibinfo {author} {\bibfnamefont {R.}~\bibnamefont
  {Gilardi}}, \bibinfo {author} {\bibfnamefont {J.}~\bibnamefont {Mesot}},
  \bibinfo {author} {\bibfnamefont {A.}~\bibnamefont {Drew}}, \bibinfo {author}
  {\bibfnamefont {U.}~\bibnamefont {Divakar}}, \bibinfo {author} {\bibfnamefont
  {S.~L.}\ \bibnamefont {Lee}}, \bibinfo {author} {\bibfnamefont {N.~H.}\
  \bibnamefont {Andersen}}, \bibinfo {author} {\bibfnamefont {J.}~\bibnamefont
  {Kohlbrecher}}, \bibinfo {author} {\bibfnamefont {N.}~\bibnamefont {Momono}},
  \ and\ \bibinfo {author} {\bibfnamefont {M.}~\bibnamefont {Oda}},\
  }\href@noop {} {\bibfield  {journal} {\bibinfo  {journal} {Physica C},\
  }\textbf {\bibinfo {volume} {408-410}},\ \bibinfo {pages} {491} (\bibinfo
  {year} {2004}{\natexlab{a}})}\BibitemShut {NoStop}%
\bibitem [{\citenamefont {Chang}\ \emph
  {et~al.}(2008){\natexlab{a}}\citenamefont {Chang}, \citenamefont
  {Niedermayer}, \citenamefont {Gilardi}, \citenamefont {Christensen},
  \citenamefont {R{\o}nnow}, \citenamefont {McMorrow}, \citenamefont {Ay},
  \citenamefont {Stahn}, \citenamefont {Sobolev}, \citenamefont {Hiess},
  \citenamefont {Pailhes}, \citenamefont {Baines}, \citenamefont {Momono},
  \citenamefont {Oda}, \citenamefont {Ido},\ and\ \citenamefont
  {Mesot}}]{Cha08}%
  \BibitemOpen
  \bibfield  {author} {\bibinfo {author} {\bibfnamefont {J.}~\bibnamefont
  {Chang}}, \bibinfo {author} {\bibfnamefont {C.}~\bibnamefont {Niedermayer}},
  \bibinfo {author} {\bibfnamefont {R.}~\bibnamefont {Gilardi}}, \bibinfo
  {author} {\bibfnamefont {N.~B.}\ \bibnamefont {Christensen}}, \bibinfo
  {author} {\bibfnamefont {H.~M.}\ \bibnamefont {R{\o}nnow}}, \bibinfo {author}
  {\bibfnamefont {D.~F.}\ \bibnamefont {McMorrow}}, \bibinfo {author}
  {\bibfnamefont {M.}~\bibnamefont {Ay}}, \bibinfo {author} {\bibfnamefont
  {J.}~\bibnamefont {Stahn}}, \bibinfo {author} {\bibfnamefont
  {O.}~\bibnamefont {Sobolev}}, \bibinfo {author} {\bibfnamefont
  {A.}~\bibnamefont {Hiess}}, \bibinfo {author} {\bibfnamefont
  {S.}~\bibnamefont {Pailhes}}, \bibinfo {author} {\bibfnamefont
  {C.}~\bibnamefont {Baines}}, \bibinfo {author} {\bibfnamefont
  {N.}~\bibnamefont {Momono}}, \bibinfo {author} {\bibfnamefont
  {M.}~\bibnamefont {Oda}}, \bibinfo {author} {\bibfnamefont {M.}~\bibnamefont
  {Ido}}, \ and\ \bibinfo {author} {\bibfnamefont {J.}~\bibnamefont {Mesot}},\
  }\href@noop {} {\bibfield  {journal} {\bibinfo  {journal} {Phys. Rev. B},\
  }\textbf {\bibinfo {volume} {78}},\ \bibinfo {pages} {104525} (\bibinfo
  {year} {2008}{\natexlab{a}})}\BibitemShut {NoStop}%
\bibitem [{\citenamefont {Khaykovich}\ \emph {et~al.}(2005)\citenamefont
  {Khaykovich}, \citenamefont {Wakimoto}, \citenamefont {Birgeneau},
  \citenamefont {Kastner}, \citenamefont {Lee}, \citenamefont {Smeibidl},
  \citenamefont {Vorderwisch},\ and\ \citenamefont {Yamada}}]{Kha05}%
  \BibitemOpen
  \bibfield  {author} {\bibinfo {author} {\bibfnamefont {B.}~\bibnamefont
  {Khaykovich}}, \bibinfo {author} {\bibfnamefont {S.}~\bibnamefont
  {Wakimoto}}, \bibinfo {author} {\bibfnamefont {R.~J.}\ \bibnamefont
  {Birgeneau}}, \bibinfo {author} {\bibfnamefont {M.~A.}\ \bibnamefont
  {Kastner}}, \bibinfo {author} {\bibfnamefont {Y.~S.}\ \bibnamefont {Lee}},
  \bibinfo {author} {\bibfnamefont {P.}~\bibnamefont {Smeibidl}}, \bibinfo
  {author} {\bibfnamefont {P.}~\bibnamefont {Vorderwisch}}, \ and\ \bibinfo
  {author} {\bibfnamefont {K.}~\bibnamefont {Yamada}},\ }\href@noop {}
  {\bibfield  {journal} {\bibinfo  {journal} {Phys. Rev. B},\ }\textbf
  {\bibinfo {volume} {71}},\ \bibinfo {pages} {220508(R)} (\bibinfo {year}
  {2005})}\BibitemShut {NoStop}%
\bibitem [{\citenamefont {Kofu}\ \emph {et~al.}(2009)\citenamefont {Kofu},
  \citenamefont {Lee}, \citenamefont {Fujita}, \citenamefont {Kang},
  \citenamefont {Eisaki},\ and\ \citenamefont {Yamada}}]{Kof09}%
  \BibitemOpen
  \bibfield  {author} {\bibinfo {author} {\bibfnamefont {M.}~\bibnamefont
  {Kofu}}, \bibinfo {author} {\bibfnamefont {S.-H.}\ \bibnamefont {Lee}},
  \bibinfo {author} {\bibfnamefont {M.}~\bibnamefont {Fujita}}, \bibinfo
  {author} {\bibfnamefont {H.-J.}\ \bibnamefont {Kang}}, \bibinfo {author}
  {\bibfnamefont {H.}~\bibnamefont {Eisaki}}, \ and\ \bibinfo {author}
  {\bibfnamefont {K.}~\bibnamefont {Yamada}},\ }\href@noop {} {\bibfield
  {journal} {\bibinfo  {journal} {Phys. Rev. Lett.},\ }\textbf {\bibinfo
  {volume} {102}},\ \bibinfo {pages} {047001} (\bibinfo {year}
  {2009})}\BibitemShut {NoStop}%
\bibitem [{\citenamefont {Haug}\ \emph {et~al.}(2010)\citenamefont {Haug},
  \citenamefont {Hinkov}, \citenamefont {Sidis}, \citenamefont {Bourges},
  \citenamefont {Christensen}, \citenamefont {Ivanov}, \citenamefont {Keller},
  \citenamefont {Lin},\ and\ \citenamefont {Keimer}}]{Hau10}%
  \BibitemOpen
  \bibfield  {author} {\bibinfo {author} {\bibfnamefont {D.}~\bibnamefont
  {Haug}}, \bibinfo {author} {\bibfnamefont {V.}~\bibnamefont {Hinkov}},
  \bibinfo {author} {\bibfnamefont {Y.}~\bibnamefont {Sidis}}, \bibinfo
  {author} {\bibfnamefont {P.}~\bibnamefont {Bourges}}, \bibinfo {author}
  {\bibfnamefont {N.~B.}\ \bibnamefont {Christensen}}, \bibinfo {author}
  {\bibfnamefont {A.}~\bibnamefont {Ivanov}}, \bibinfo {author} {\bibfnamefont
  {T.}~\bibnamefont {Keller}}, \bibinfo {author} {\bibfnamefont {C.~T.}\
  \bibnamefont {Lin}}, \ and\ \bibinfo {author} {\bibfnamefont
  {B.}~\bibnamefont {Keimer}},\ }\href@noop {} {\bibfield  {journal} {\bibinfo
  {journal} {New J. Phys.},\ }\textbf {\bibinfo {volume} {12}},\ \bibinfo
  {pages} {105006} (\bibinfo {year} {2010})}\BibitemShut {NoStop}%
\bibitem [{\citenamefont {Chang}\ \emph {et~al.}(2006)\citenamefont {Chang},
  \citenamefont {Mesot}, \citenamefont {Gilardi}, \citenamefont {Kohlbrecher},
  \citenamefont {Drew}, \citenamefont {Divakar}, \citenamefont {Lister},
  \citenamefont {Lee}, \citenamefont {Brown}, \citenamefont {Charalambous},
  \citenamefont {Forgan}, \citenamefont {Dewhurst}, \citenamefont {Cubitt},
  \citenamefont {Momono},\ and\ \citenamefont {Oda}}]{Cha06}%
  \BibitemOpen
  \bibfield  {author} {\bibinfo {author} {\bibfnamefont {J.}~\bibnamefont
  {Chang}}, \bibinfo {author} {\bibfnamefont {J.}~\bibnamefont {Mesot}},
  \bibinfo {author} {\bibfnamefont {R.}~\bibnamefont {Gilardi}}, \bibinfo
  {author} {\bibfnamefont {J.}~\bibnamefont {Kohlbrecher}}, \bibinfo {author}
  {\bibfnamefont {A.~J.}\ \bibnamefont {Drew}}, \bibinfo {author}
  {\bibfnamefont {U.}~\bibnamefont {Divakar}}, \bibinfo {author} {\bibfnamefont
  {S.~J.}\ \bibnamefont {Lister}}, \bibinfo {author} {\bibfnamefont {S.~L.}\
  \bibnamefont {Lee}}, \bibinfo {author} {\bibfnamefont {S.~P.}\ \bibnamefont
  {Brown}}, \bibinfo {author} {\bibfnamefont {D.}~\bibnamefont {Charalambous}},
  \bibinfo {author} {\bibfnamefont {E.~M.}\ \bibnamefont {Forgan}}, \bibinfo
  {author} {\bibfnamefont {C.~D.}\ \bibnamefont {Dewhurst}}, \bibinfo {author}
  {\bibfnamefont {R.}~\bibnamefont {Cubitt}}, \bibinfo {author} {\bibfnamefont
  {N.}~\bibnamefont {Momono}}, \ and\ \bibinfo {author} {\bibfnamefont
  {M.}~\bibnamefont {Oda}},\ }\href@noop {} {\bibfield  {journal} {\bibinfo
  {journal} {Physica B},\ }\textbf {\bibinfo {volume} {385-386}},\ \bibinfo
  {pages} {35} (\bibinfo {year} {2006})}\BibitemShut {NoStop}%
\bibitem [{\citenamefont {Nakano}\ \emph {et~al.}(1998)\citenamefont {Nakano},
  \citenamefont {Momono}, \citenamefont {Oda},\ and\ \citenamefont
  {Ido}}]{Nak98}%
  \BibitemOpen
  \bibfield  {author} {\bibinfo {author} {\bibfnamefont {T.}~\bibnamefont
  {Nakano}}, \bibinfo {author} {\bibfnamefont {N.}~\bibnamefont {Momono}},
  \bibinfo {author} {\bibfnamefont {M.}~\bibnamefont {Oda}}, \ and\ \bibinfo
  {author} {\bibfnamefont {M.}~\bibnamefont {Ido}},\ }\href@noop {} {\bibfield
  {journal} {\bibinfo  {journal} {J. Phys. Soc. Jpn.},\ }\textbf {\bibinfo
  {volume} {67}},\ \bibinfo {pages} {2622} (\bibinfo {year}
  {1998})}\BibitemShut {NoStop}%
\bibitem [{\citenamefont {Chang}\ \emph {et~al.}(2009)\citenamefont {Chang},
  \citenamefont {Christensen}, \citenamefont {Niedermayer}, \citenamefont
  {Lefmann}, \citenamefont {R\o~nnow}, \citenamefont {McMorrow}, \citenamefont
  {Schneidewind}, \citenamefont {Link}, \citenamefont {Hiess}, \citenamefont
  {Boehm}, \citenamefont {Mottl}, \citenamefont {Pailh\'{e}s}, \citenamefont
  {Momono}, \citenamefont {Oda}, \citenamefont {Ido},\ and\ \citenamefont
  {J.}}]{Cha09}%
  \BibitemOpen
  \bibfield  {author} {\bibinfo {author} {\bibfnamefont {J.}~\bibnamefont
  {Chang}}, \bibinfo {author} {\bibfnamefont {N.~B.}\ \bibnamefont
  {Christensen}}, \bibinfo {author} {\bibfnamefont {C.}~\bibnamefont
  {Niedermayer}}, \bibinfo {author} {\bibfnamefont {K.}~\bibnamefont
  {Lefmann}}, \bibinfo {author} {\bibfnamefont {H.~M.}\ \bibnamefont
  {R\o~nnow}}, \bibinfo {author} {\bibfnamefont {D.~F.}\ \bibnamefont
  {McMorrow}}, \bibinfo {author} {\bibfnamefont {A.}~\bibnamefont
  {Schneidewind}}, \bibinfo {author} {\bibfnamefont {P.}~\bibnamefont {Link}},
  \bibinfo {author} {\bibfnamefont {A.}~\bibnamefont {Hiess}}, \bibinfo
  {author} {\bibfnamefont {M.}~\bibnamefont {Boehm}}, \bibinfo {author}
  {\bibfnamefont {R.}~\bibnamefont {Mottl}}, \bibinfo {author} {\bibfnamefont
  {S.}~\bibnamefont {Pailh\'{e}s}}, \bibinfo {author} {\bibfnamefont
  {N.}~\bibnamefont {Momono}}, \bibinfo {author} {\bibfnamefont
  {M.}~\bibnamefont {Oda}}, \bibinfo {author} {\bibfnamefont {M.}~\bibnamefont
  {Ido}}, \ and\ \bibinfo {author} {\bibfnamefont {M.}~\bibnamefont {J.}},\
  }\href@noop {} {\bibfield  {journal} {\bibinfo  {journal} {Phys. Rev.
  Lett.},\ }\textbf {\bibinfo {volume} {102}},\ \bibinfo {pages} {177006}
  (\bibinfo {year} {2009})}\BibitemShut {NoStop}%
\bibitem [{\citenamefont {Lipscombe}\ \emph {et~al.}(2007)\citenamefont
  {Lipscombe}, \citenamefont {Hayden}, \citenamefont {Vignolle}, \citenamefont
  {McMorrow},\ and\ \citenamefont {Perring}}]{Lip07}%
  \BibitemOpen
  \bibfield  {author} {\bibinfo {author} {\bibfnamefont {O.~J.}\ \bibnamefont
  {Lipscombe}}, \bibinfo {author} {\bibfnamefont {S.~M.}\ \bibnamefont
  {Hayden}}, \bibinfo {author} {\bibfnamefont {B.}~\bibnamefont {Vignolle}},
  \bibinfo {author} {\bibfnamefont {D.~F.}\ \bibnamefont {McMorrow}}, \ and\
  \bibinfo {author} {\bibfnamefont {T.~G.}\ \bibnamefont {Perring}},\
  }\href@noop {} {\bibfield  {journal} {\bibinfo  {journal} {Phys. Rev.
  Lett.},\ }\textbf {\bibinfo {volume} {99}},\ \bibinfo {pages} {067002}
  (\bibinfo {year} {2007})}\BibitemShut {NoStop}%
\bibitem [{\citenamefont {Khaykovich}\ \emph {et~al.}(2002)\citenamefont
  {Khaykovich}, \citenamefont {Lee}, \citenamefont {Erwin}, \citenamefont
  {Lee}, \citenamefont {Wakimoto}, \citenamefont {Thomas}, \citenamefont
  {Kastner},\ and\ \citenamefont {Birgeneau}}]{Kha02}%
  \BibitemOpen
  \bibfield  {author} {\bibinfo {author} {\bibfnamefont {B.}~\bibnamefont
  {Khaykovich}}, \bibinfo {author} {\bibfnamefont {Y.~S.}\ \bibnamefont {Lee}},
  \bibinfo {author} {\bibfnamefont {R.~W.}\ \bibnamefont {Erwin}}, \bibinfo
  {author} {\bibfnamefont {S.-H.}\ \bibnamefont {Lee}}, \bibinfo {author}
  {\bibfnamefont {S.}~\bibnamefont {Wakimoto}}, \bibinfo {author}
  {\bibfnamefont {K.~J.}\ \bibnamefont {Thomas}}, \bibinfo {author}
  {\bibfnamefont {M.~A.}\ \bibnamefont {Kastner}}, \ and\ \bibinfo {author}
  {\bibfnamefont {R.~J.}\ \bibnamefont {Birgeneau}},\ }\href@noop {} {\bibfield
   {journal} {\bibinfo  {journal} {Phys. Rev. B},\ }\textbf {\bibinfo {volume}
  {66}},\ \bibinfo {pages} {014528} (\bibinfo {year} {2002})}\BibitemShut
  {NoStop}%
\bibitem [{\citenamefont {Khaykovich}\ \emph {et~al.}(2003)\citenamefont
  {Khaykovich}, \citenamefont {Birgeneau}, \citenamefont {Chou}, \citenamefont
  {Erwin}, \citenamefont {Kastner}, \citenamefont {Lee}, \citenamefont {Lee},
  \citenamefont {Smeibidl}, \citenamefont {Vorderwisch},\ and\ \citenamefont
  {Wakimoto}}]{Kha03}%
  \BibitemOpen
  \bibfield  {author} {\bibinfo {author} {\bibfnamefont {B.}~\bibnamefont
  {Khaykovich}}, \bibinfo {author} {\bibfnamefont {R.~J.}\ \bibnamefont
  {Birgeneau}}, \bibinfo {author} {\bibfnamefont {F.~C.}\ \bibnamefont {Chou}},
  \bibinfo {author} {\bibfnamefont {R.~W.}\ \bibnamefont {Erwin}}, \bibinfo
  {author} {\bibfnamefont {M.~A.}\ \bibnamefont {Kastner}}, \bibinfo {author}
  {\bibfnamefont {S.-H.}\ \bibnamefont {Lee}}, \bibinfo {author} {\bibfnamefont
  {Y.~S.}\ \bibnamefont {Lee}}, \bibinfo {author} {\bibfnamefont
  {P.}~\bibnamefont {Smeibidl}}, \bibinfo {author} {\bibfnamefont
  {P.}~\bibnamefont {Vorderwisch}}, \ and\ \bibinfo {author} {\bibfnamefont
  {S.}~\bibnamefont {Wakimoto}},\ }\href@noop {} {\bibfield  {journal}
  {\bibinfo  {journal} {Phys. Rev. B},\ }\textbf {\bibinfo {volume} {67}},\
  \bibinfo {pages} {054501} (\bibinfo {year} {2003})}\BibitemShut {NoStop}%
\bibitem [{\citenamefont {Katano}\ \emph {et~al.}(2000)\citenamefont {Katano},
  \citenamefont {Sato}, \citenamefont {Yamada}, \citenamefont {Suzuki},\ and\
  \citenamefont {Fukase}}]{Kat00}%
  \BibitemOpen
  \bibfield  {author} {\bibinfo {author} {\bibfnamefont {S.}~\bibnamefont
  {Katano}}, \bibinfo {author} {\bibfnamefont {M.}~\bibnamefont {Sato}},
  \bibinfo {author} {\bibfnamefont {K.}~\bibnamefont {Yamada}}, \bibinfo
  {author} {\bibfnamefont {T.}~\bibnamefont {Suzuki}}, \ and\ \bibinfo {author}
  {\bibfnamefont {T.}~\bibnamefont {Fukase}},\ }\href@noop {} {\bibfield
  {journal} {\bibinfo  {journal} {Phys. Rev. B},\ }\textbf {\bibinfo {volume}
  {62}},\ \bibinfo {pages} {R14677} (\bibinfo {year} {2000})}\BibitemShut
  {NoStop}%
\bibitem [{\citenamefont {Kohlbrecher}\ and\ \citenamefont
  {Wagner}(2000)}]{Koh00}%
  \BibitemOpen
  \bibfield  {author} {\bibinfo {author} {\bibfnamefont {J.}~\bibnamefont
  {Kohlbrecher}}\ and\ \bibinfo {author} {\bibfnamefont {W.}~\bibnamefont
  {Wagner}},\ }\href@noop {} {\bibfield  {journal} {\bibinfo  {journal} {J.
  Appl. Cryst.},\ }\textbf {\bibinfo {volume} {33}},\ \bibinfo {pages} {804}
  (\bibinfo {year} {2000})}\BibitemShut {NoStop}%
\bibitem [{\citenamefont {Gilardi}\ \emph
  {et~al.}(2004){\natexlab{b}}\citenamefont {Gilardi}, \citenamefont {Mesot},
  \citenamefont {Brown}, \citenamefont {Forgan}, \citenamefont {Drew},
  \citenamefont {Lee}, \citenamefont {Cubitt}, \citenamefont {Dewhurst},
  \citenamefont {Uefuji},\ and\ \citenamefont {Yamada}}]{Gil04b}%
  \BibitemOpen
  \bibfield  {author} {\bibinfo {author} {\bibfnamefont {R.}~\bibnamefont
  {Gilardi}}, \bibinfo {author} {\bibfnamefont {J.}~\bibnamefont {Mesot}},
  \bibinfo {author} {\bibfnamefont {S.~P.}\ \bibnamefont {Brown}}, \bibinfo
  {author} {\bibfnamefont {S.~P.}\ \bibnamefont {Forgan}}, \bibinfo {author}
  {\bibfnamefont {A.}~\bibnamefont {Drew}}, \bibinfo {author} {\bibfnamefont
  {S.~L.}\ \bibnamefont {Lee}}, \bibinfo {author} {\bibfnamefont
  {R.}~\bibnamefont {Cubitt}}, \bibinfo {author} {\bibfnamefont {C.~D.}\
  \bibnamefont {Dewhurst}}, \bibinfo {author} {\bibfnamefont {T.}~\bibnamefont
  {Uefuji}}, \ and\ \bibinfo {author} {\bibfnamefont {K.}~\bibnamefont
  {Yamada}},\ }\href@noop {} {\bibfield  {journal} {\bibinfo  {journal} {Phys.
  Rev. Lett.},\ }\textbf {\bibinfo {volume} {93}},\ \bibinfo {pages} {217001}
  (\bibinfo {year} {2004}{\natexlab{b}})}\BibitemShut {NoStop}%
\bibitem [{\citenamefont {Chang}\ and\ \citenamefont {Mesot}(2008)}]{Cha08a}%
  \BibitemOpen
  \bibfield  {author} {\bibinfo {author} {\bibfnamefont {J.}~\bibnamefont
  {Chang}}\ and\ \bibinfo {author} {\bibfnamefont {J.}~\bibnamefont {Mesot}},\
  }\href@noop {} {\bibfield  {journal} {\bibinfo  {journal} {PRAMANA-J.
  Phys.},\ }\textbf {\bibinfo {volume} {71}},\ \bibinfo {pages} {679} (\bibinfo
  {year} {2008})}\BibitemShut {NoStop}%
\bibitem [{\citenamefont {Nakai}\ \emph {et~al.}(2002)\citenamefont {Nakai},
  \citenamefont {Miranovi\'{c}}, \citenamefont {Ichioka},\ and\ \citenamefont
  {Machida}}]{Nak02}%
  \BibitemOpen
  \bibfield  {author} {\bibinfo {author} {\bibfnamefont {N.}~\bibnamefont
  {Nakai}}, \bibinfo {author} {\bibfnamefont {P.}~\bibnamefont
  {Miranovi\'{c}}}, \bibinfo {author} {\bibfnamefont {M.}~\bibnamefont
  {Ichioka}}, \ and\ \bibinfo {author} {\bibfnamefont {K.}~\bibnamefont
  {Machida}},\ }\href@noop {} {\bibfield  {journal} {\bibinfo  {journal} {Phys.
  Rev. Lett.},\ }\textbf {\bibinfo {volume} {89}},\ \bibinfo {pages} {237004}
  (\bibinfo {year} {2002})}\BibitemShut {NoStop}%
\bibitem [{\citenamefont {Keimer}\ \emph {et~al.}(1994)\citenamefont {Keimer},
  \citenamefont {Shih}, \citenamefont {Erwin}, \citenamefont {Lynn},
  \citenamefont {Dogan},\ and\ \citenamefont {Aksay}}]{Kei94}%
  \BibitemOpen
  \bibfield  {author} {\bibinfo {author} {\bibfnamefont {B.}~\bibnamefont
  {Keimer}}, \bibinfo {author} {\bibfnamefont {W.~Y.}\ \bibnamefont {Shih}},
  \bibinfo {author} {\bibfnamefont {R.~W.}\ \bibnamefont {Erwin}}, \bibinfo
  {author} {\bibfnamefont {J.~W.}\ \bibnamefont {Lynn}}, \bibinfo {author}
  {\bibfnamefont {F.}~\bibnamefont {Dogan}}, \ and\ \bibinfo {author}
  {\bibfnamefont {I.~A.}\ \bibnamefont {Aksay}},\ }\href@noop {} {\bibfield
  {journal} {\bibinfo  {journal} {Phys. Rev. Lett.},\ }\textbf {\bibinfo
  {volume} {73}},\ \bibinfo {pages} {3459} (\bibinfo {year}
  {1994})}\BibitemShut {NoStop}%
\bibitem [{\citenamefont {Brown}\ \emph {et~al.}(2004)\citenamefont {Brown},
  \citenamefont {Charalambous}, \citenamefont {Jones}, \citenamefont {Forgan},
  \citenamefont {Kealey}, \citenamefont {Erb},\ and\ \citenamefont
  {Kohlbrecher}}]{Bro04}%
  \BibitemOpen
  \bibfield  {author} {\bibinfo {author} {\bibfnamefont {S.~P.}\ \bibnamefont
  {Brown}}, \bibinfo {author} {\bibfnamefont {D.}~\bibnamefont {Charalambous}},
  \bibinfo {author} {\bibfnamefont {E.~C.}\ \bibnamefont {Jones}}, \bibinfo
  {author} {\bibfnamefont {E.~M.}\ \bibnamefont {Forgan}}, \bibinfo {author}
  {\bibfnamefont {P.~G.}\ \bibnamefont {Kealey}}, \bibinfo {author}
  {\bibfnamefont {A.}~\bibnamefont {Erb}}, \ and\ \bibinfo {author}
  {\bibfnamefont {J.}~\bibnamefont {Kohlbrecher}},\ }\href@noop {} {\bibfield
  {journal} {\bibinfo  {journal} {Phys. Rev. Lett.},\ }\textbf {\bibinfo
  {volume} {92}},\ \bibinfo {pages} {067004} (\bibinfo {year}
  {2004})}\BibitemShut {NoStop}%
\bibitem [{\citenamefont {Eskildsen}\ \emph {et~al.}(2003)\citenamefont
  {Eskildsen}, \citenamefont {Dewhurst}, \citenamefont {Hoogenboom},
  \citenamefont {Petrovic},\ and\ \citenamefont {Canfield}}]{Esk03}%
  \BibitemOpen
  \bibfield  {author} {\bibinfo {author} {\bibfnamefont {M.~R.}\ \bibnamefont
  {Eskildsen}}, \bibinfo {author} {\bibfnamefont {C.~D.}\ \bibnamefont
  {Dewhurst}}, \bibinfo {author} {\bibfnamefont {B.~W.}\ \bibnamefont
  {Hoogenboom}}, \bibinfo {author} {\bibfnamefont {C.}~\bibnamefont
  {Petrovic}}, \ and\ \bibinfo {author} {\bibfnamefont {P.~C.}\ \bibnamefont
  {Canfield}},\ }\href@noop {} {\bibfield  {journal} {\bibinfo  {journal}
  {Phys. Rev. Lett.},\ }\textbf {\bibinfo {volume} {90}},\ \bibinfo {pages}
  {187001} (\bibinfo {year} {2003})}\BibitemShut {NoStop}%
\bibitem [{\citenamefont {Chakravarty}\ \emph {et~al.}(1993)\citenamefont
  {Chakravarty}, \citenamefont {Sudb\o{}}, \citenamefont {Anderson},\ and\
  \citenamefont {Strong}}]{Cha93}%
  \BibitemOpen
  \bibfield  {author} {\bibinfo {author} {\bibfnamefont {S.}~\bibnamefont
  {Chakravarty}}, \bibinfo {author} {\bibfnamefont {A.}~\bibnamefont
  {Sudb\o{}}}, \bibinfo {author} {\bibfnamefont {P.~W.}\ \bibnamefont
  {Anderson}}, \ and\ \bibinfo {author} {\bibfnamefont {S.}~\bibnamefont
  {Strong}},\ }\href@noop {} {\bibfield  {journal} {\bibinfo  {journal}
  {Science},\ }\textbf {\bibinfo {volume} {261}},\ \bibinfo {pages} {337}
  (\bibinfo {year} {1993})}\BibitemShut {NoStop}%
\bibitem [{\citenamefont {Yoshida}\ \emph {et~al.}(2006)\citenamefont
  {Yoshida}, \citenamefont {Zhou}, \citenamefont {Tanaka}, \citenamefont
  {Yang}, \citenamefont {Hussain}, \citenamefont {Shen}, \citenamefont
  {Fujimori}, \citenamefont {Sahrakorpi}, \citenamefont {Lindroos},
  \citenamefont {Markiewicz}, \citenamefont {Bansil}, \citenamefont {Komiya},
  \citenamefont {Ando}, \citenamefont {Eisaki}, \citenamefont {Kakeshita},\
  and\ \citenamefont {Uchida}}]{Yos06}%
  \BibitemOpen
  \bibfield  {author} {\bibinfo {author} {\bibfnamefont {T.}~\bibnamefont
  {Yoshida}}, \bibinfo {author} {\bibfnamefont {X.~J.}\ \bibnamefont {Zhou}},
  \bibinfo {author} {\bibfnamefont {K.}~\bibnamefont {Tanaka}}, \bibinfo
  {author} {\bibfnamefont {W.~L.}\ \bibnamefont {Yang}}, \bibinfo {author}
  {\bibfnamefont {Z.}~\bibnamefont {Hussain}}, \bibinfo {author} {\bibfnamefont
  {Z.-X.}\ \bibnamefont {Shen}}, \bibinfo {author} {\bibfnamefont
  {A.}~\bibnamefont {Fujimori}}, \bibinfo {author} {\bibfnamefont
  {S.}~\bibnamefont {Sahrakorpi}}, \bibinfo {author} {\bibfnamefont
  {M.}~\bibnamefont {Lindroos}}, \bibinfo {author} {\bibfnamefont {R.~S.}\
  \bibnamefont {Markiewicz}}, \bibinfo {author} {\bibfnamefont
  {A.}~\bibnamefont {Bansil}}, \bibinfo {author} {\bibfnamefont
  {S.}~\bibnamefont {Komiya}}, \bibinfo {author} {\bibfnamefont
  {Y.}~\bibnamefont {Ando}}, \bibinfo {author} {\bibfnamefont {H.}~\bibnamefont
  {Eisaki}}, \bibinfo {author} {\bibfnamefont {T.}~\bibnamefont {Kakeshita}}, \
  and\ \bibinfo {author} {\bibfnamefont {S.}~\bibnamefont {Uchida}},\
  }\href@noop {} {\bibfield  {journal} {\bibinfo  {journal} {Phys. Rev. B},\
  }\textbf {\bibinfo {volume} {74}},\ \bibinfo {pages} {224510} (\bibinfo
  {year} {2006})}\BibitemShut {NoStop}%
\bibitem [{\citenamefont {Chang}\ \emph
  {et~al.}(2008){\natexlab{b}}\citenamefont {Chang}, \citenamefont {Shi},
  \citenamefont {Pailh\'es}, \citenamefont {M\aa{}nsson}, \citenamefont
  {Claesson}, \citenamefont {Tjernberg}, \citenamefont {Bendounan},
  \citenamefont {Sassa}, \citenamefont {Patthey}, \citenamefont {Momono},
  \citenamefont {Oda}, \citenamefont {Ido}, \citenamefont {Guerrero},
  \citenamefont {Mudry},\ and\ \citenamefont {Mesot}}]{Cha08b}%
  \BibitemOpen
  \bibfield  {author} {\bibinfo {author} {\bibfnamefont {J.}~\bibnamefont
  {Chang}}, \bibinfo {author} {\bibfnamefont {M.}~\bibnamefont {Shi}}, \bibinfo
  {author} {\bibfnamefont {S.}~\bibnamefont {Pailh\'es}}, \bibinfo {author}
  {\bibfnamefont {M.}~\bibnamefont {M\aa{}nsson}}, \bibinfo {author}
  {\bibfnamefont {T.}~\bibnamefont {Claesson}}, \bibinfo {author}
  {\bibfnamefont {O.}~\bibnamefont {Tjernberg}}, \bibinfo {author}
  {\bibfnamefont {A.}~\bibnamefont {Bendounan}}, \bibinfo {author}
  {\bibfnamefont {Y.}~\bibnamefont {Sassa}}, \bibinfo {author} {\bibfnamefont
  {L.}~\bibnamefont {Patthey}}, \bibinfo {author} {\bibfnamefont
  {N.}~\bibnamefont {Momono}}, \bibinfo {author} {\bibfnamefont
  {M.}~\bibnamefont {Oda}}, \bibinfo {author} {\bibfnamefont {M.}~\bibnamefont
  {Ido}}, \bibinfo {author} {\bibfnamefont {S.}~\bibnamefont {Guerrero}},
  \bibinfo {author} {\bibfnamefont {C.}~\bibnamefont {Mudry}}, \ and\ \bibinfo
  {author} {\bibfnamefont {J.}~\bibnamefont {Mesot}},\ }\href@noop {}
  {\bibfield  {journal} {\bibinfo  {journal} {Phys. Rev. B},\ }\textbf
  {\bibinfo {volume} {78}},\ \bibinfo {pages} {205103} (\bibinfo {year}
  {2008}{\natexlab{b}})}\BibitemShut {NoStop}%
\bibitem [{\citenamefont {White}\ \emph {et~al.}(2008)\citenamefont {White},
  \citenamefont {Brown}, \citenamefont {Forgan}, \citenamefont {Laver},
  \citenamefont {Bowell}, \citenamefont {Lycett}, \citenamefont {Charalambous},
  \citenamefont {Hinkov}, \citenamefont {Erb},\ and\ \citenamefont
  {Kohlbrecher}}]{Whi08}%
  \BibitemOpen
  \bibfield  {author} {\bibinfo {author} {\bibfnamefont {J.~S.}\ \bibnamefont
  {White}}, \bibinfo {author} {\bibfnamefont {S.~P.}\ \bibnamefont {Brown}},
  \bibinfo {author} {\bibfnamefont {E.~M.}\ \bibnamefont {Forgan}}, \bibinfo
  {author} {\bibfnamefont {M.}~\bibnamefont {Laver}}, \bibinfo {author}
  {\bibfnamefont {C.~J.}\ \bibnamefont {Bowell}}, \bibinfo {author}
  {\bibfnamefont {R.~J.}\ \bibnamefont {Lycett}}, \bibinfo {author}
  {\bibfnamefont {D.}~\bibnamefont {Charalambous}}, \bibinfo {author}
  {\bibfnamefont {V.}~\bibnamefont {Hinkov}}, \bibinfo {author} {\bibfnamefont
  {A.}~\bibnamefont {Erb}}, \ and\ \bibinfo {author} {\bibfnamefont
  {J.}~\bibnamefont {Kohlbrecher}},\ }\href@noop {} {\bibfield  {journal}
  {\bibinfo  {journal} {Phys. Rev. B},\ }\textbf {\bibinfo {volume} {78}},\
  \bibinfo {pages} {174513} (\bibinfo {year} {2008})}\BibitemShut {NoStop}%
\bibitem [{\citenamefont {White}\ \emph {et~al.}(2009)\citenamefont {White},
  \citenamefont {Hinkov}, \citenamefont {Heslop}, \citenamefont {Lycett},
  \citenamefont {Forgan}, \citenamefont {Bowell}, \citenamefont {Str\"{a}ssle},
  \citenamefont {Abrahamsen}, \citenamefont {Laver}, \citenamefont {Dewhurst},
  \citenamefont {Kohlbrecher}, \citenamefont {Gavilano}, \citenamefont {Mesot},
  \citenamefont {Keimer},\ and\ \citenamefont {Erb}}]{Whi09}%
  \BibitemOpen
  \bibfield  {author} {\bibinfo {author} {\bibfnamefont {J.~S.}\ \bibnamefont
  {White}}, \bibinfo {author} {\bibfnamefont {V.}~\bibnamefont {Hinkov}},
  \bibinfo {author} {\bibfnamefont {R.~W.}\ \bibnamefont {Heslop}}, \bibinfo
  {author} {\bibfnamefont {R.~J.}\ \bibnamefont {Lycett}}, \bibinfo {author}
  {\bibfnamefont {E.~M.}\ \bibnamefont {Forgan}}, \bibinfo {author}
  {\bibfnamefont {C.}~\bibnamefont {Bowell}}, \bibinfo {author} {\bibfnamefont
  {S.}~\bibnamefont {Str\"{a}ssle}}, \bibinfo {author} {\bibfnamefont {A.~B.}\
  \bibnamefont {Abrahamsen}}, \bibinfo {author} {\bibfnamefont
  {M.}~\bibnamefont {Laver}}, \bibinfo {author} {\bibfnamefont {C.~D.}\
  \bibnamefont {Dewhurst}}, \bibinfo {author} {\bibfnamefont {J.}~\bibnamefont
  {Kohlbrecher}}, \bibinfo {author} {\bibfnamefont {J.~L.}\ \bibnamefont
  {Gavilano}}, \bibinfo {author} {\bibfnamefont {J.}~\bibnamefont {Mesot}},
  \bibinfo {author} {\bibfnamefont {B.}~\bibnamefont {Keimer}}, \ and\ \bibinfo
  {author} {\bibfnamefont {A.}~\bibnamefont {Erb}},\ }\href@noop {} {\bibfield
  {journal} {\bibinfo  {journal} {Phys. Rev. Lett.},\ }\textbf {\bibinfo
  {volume} {102}},\ \bibinfo {pages} {097001} (\bibinfo {year}
  {2009})}\BibitemShut {NoStop}%
\bibitem [{\citenamefont {White}\ \emph {et~al.}(2011)\citenamefont {White},
  \citenamefont {Heslop}, \citenamefont {Holmes}, \citenamefont {Forgan},
  \citenamefont {Hinkov}, \citenamefont {Egetenmeyer}, \citenamefont
  {Gavilano}, \citenamefont {Laver}, \citenamefont {Dewhurst}, \citenamefont
  {Cubitt},\ and\ \citenamefont {Erb}}]{Whi11}%
  \BibitemOpen
  \bibfield  {author} {\bibinfo {author} {\bibfnamefont {J.~S.}\ \bibnamefont
  {White}}, \bibinfo {author} {\bibfnamefont {R.~W.}\ \bibnamefont {Heslop}},
  \bibinfo {author} {\bibfnamefont {A.~T.}\ \bibnamefont {Holmes}}, \bibinfo
  {author} {\bibfnamefont {E.~M.}\ \bibnamefont {Forgan}}, \bibinfo {author}
  {\bibfnamefont {V.}~\bibnamefont {Hinkov}}, \bibinfo {author} {\bibfnamefont
  {N.}~\bibnamefont {Egetenmeyer}}, \bibinfo {author} {\bibfnamefont {J.~L.}\
  \bibnamefont {Gavilano}}, \bibinfo {author} {\bibfnamefont {M.}~\bibnamefont
  {Laver}}, \bibinfo {author} {\bibfnamefont {C.~D.}\ \bibnamefont {Dewhurst}},
  \bibinfo {author} {\bibfnamefont {R.}~\bibnamefont {Cubitt}}, \ and\ \bibinfo
  {author} {\bibfnamefont {A.}~\bibnamefont {Erb}},\ }\href@noop {} {\bibfield
  {journal} {\bibinfo  {journal} {Phys. Rev. B},\ }\textbf {\bibinfo {volume}
  {84}},\ \bibinfo {pages} {104519} (\bibinfo {year} {2011})}\BibitemShut
  {NoStop}%
\bibitem [{\citenamefont {Rourke}\ \emph {et~al.}(2011)\citenamefont {Rourke},
  \citenamefont {Mouzopoulou}, \citenamefont {Xu}, \citenamefont
  {Panagopoulos}, \citenamefont {Wang}, \citenamefont {Vignolle}, \citenamefont
  {Proust}, \citenamefont {Kurganova}, \citenamefont {Zeitler}, \citenamefont
  {Tanabe}, \citenamefont {Adachi}, \citenamefont {Koike},\ and\ \citenamefont
  {Hussey}}]{Rou11}%
  \BibitemOpen
  \bibfield  {author} {\bibinfo {author} {\bibfnamefont {P.~M.~C.}\
  \bibnamefont {Rourke}}, \bibinfo {author} {\bibfnamefont {I.}~\bibnamefont
  {Mouzopoulou}}, \bibinfo {author} {\bibfnamefont {X.}~\bibnamefont {Xu}},
  \bibinfo {author} {\bibfnamefont {C.}~\bibnamefont {Panagopoulos}}, \bibinfo
  {author} {\bibfnamefont {Y.}~\bibnamefont {Wang}}, \bibinfo {author}
  {\bibfnamefont {B.}~\bibnamefont {Vignolle}}, \bibinfo {author}
  {\bibfnamefont {C.}~\bibnamefont {Proust}}, \bibinfo {author} {\bibfnamefont
  {E.~V.}\ \bibnamefont {Kurganova}}, \bibinfo {author} {\bibfnamefont
  {U.}~\bibnamefont {Zeitler}}, \bibinfo {author} {\bibfnamefont
  {Y.}~\bibnamefont {Tanabe}}, \bibinfo {author} {\bibfnamefont
  {T.}~\bibnamefont {Adachi}}, \bibinfo {author} {\bibfnamefont
  {Y.}~\bibnamefont {Koike}}, \ and\ \bibinfo {author} {\bibfnamefont {N.~E.}\
  \bibnamefont {Hussey}},\ }\href@noop {} {\bibfield  {journal} {\bibinfo
  {journal} {Nature Physics},\ }\textbf {\bibinfo {volume} {7}},\ \bibinfo
  {pages} {455–458} (\bibinfo {year} {2011})}\BibitemShut {NoStop}%
\bibitem [{\citenamefont {Wang}\ and\ \citenamefont {Wen}(2008)}]{Wan08}%
  \BibitemOpen
  \bibfield  {author} {\bibinfo {author} {\bibfnamefont {Y.}~\bibnamefont
  {Wang}}\ and\ \bibinfo {author} {\bibfnamefont {H.-H.}\ \bibnamefont {Wen}},\
  }\href@noop {} {\bibfield  {journal} {\bibinfo  {journal} {Eur. Phys.
  Lett.},\ }\textbf {\bibinfo {volume} {81}},\ \bibinfo {pages} {57007}
  (\bibinfo {year} {2008})}\BibitemShut {NoStop}%
\bibitem [{\citenamefont {Schafgans}\ \emph {et~al.}(2010)\citenamefont
  {Schafgans}, \citenamefont {LaForge}, \citenamefont {Dordevic}, \citenamefont
  {Qazilbash}, \citenamefont {Padilla}, \citenamefont {Burch}, \citenamefont
  {Li}, \citenamefont {Komiya}, \citenamefont {Ando},\ and\ \citenamefont
  {Basov}}]{Sch10}%
  \BibitemOpen
  \bibfield  {author} {\bibinfo {author} {\bibfnamefont {A.~A.}\ \bibnamefont
  {Schafgans}}, \bibinfo {author} {\bibfnamefont {A.~D.}\ \bibnamefont
  {LaForge}}, \bibinfo {author} {\bibfnamefont {S.~V.}\ \bibnamefont
  {Dordevic}}, \bibinfo {author} {\bibfnamefont {M.~M.}\ \bibnamefont
  {Qazilbash}}, \bibinfo {author} {\bibfnamefont {W.~J.}\ \bibnamefont
  {Padilla}}, \bibinfo {author} {\bibfnamefont {K.~S.}\ \bibnamefont {Burch}},
  \bibinfo {author} {\bibfnamefont {Z.~Q.}\ \bibnamefont {Li}}, \bibinfo
  {author} {\bibfnamefont {S.}~\bibnamefont {Komiya}}, \bibinfo {author}
  {\bibfnamefont {Y.}~\bibnamefont {Ando}}, \ and\ \bibinfo {author}
  {\bibfnamefont {D.~N.}\ \bibnamefont {Basov}},\ }\href@noop {} {\bibfield
  {journal} {\bibinfo  {journal} {Phys. Rev. Lett.},\ }\textbf {\bibinfo
  {volume} {104}},\ \bibinfo {pages} {157002} (\bibinfo {year}
  {2010})}\BibitemShut {NoStop}%
\bibitem [{\citenamefont {Clem}(1975)}]{Cle75}%
  \BibitemOpen
  \bibfield  {author} {\bibinfo {author} {\bibfnamefont {J.~R.}\ \bibnamefont
  {Clem}},\ }\href@noop {} {\bibfield  {journal} {\bibinfo  {journal} {J. Low
  Temp. Phys.},\ }\textbf {\bibinfo {volume} {18}},\ \bibinfo {pages} {427}
  (\bibinfo {year} {1975})}\BibitemShut {NoStop}%
\bibitem [{Not(){\natexlab{a}}}]{Not01}%
  \BibitemOpen
  \href@noop {} {} ({\natexlab{a}}),\ \bibinfo {note} {in our field range
  $\mu_0 H \gtrsim 0.05$\,T, $\mathcal{F}(H)$ is practically independent of the
  superconducting penetration depth $\lambda$. In the Clem model,\citep{Cle75}
  the argument $z$ of the modified Bessel function $K_1(z)$ is strictly $z =
  (G^2 + \lambda^{-2})^{1/2} \xi$. Here $G \gtrsim 0.003$\,\AA$^{-1}$ and
  $1/\lambda \lesssim 0.0005$\,\AA$^{-1}$,\citep{Pan99} so $z \simeq G \xi$. In
  the Clem model, $\lambda$ also enters as an $H$-independent prefactor
  $1/\lambda K_1(\xi/\lambda)$.}\BibitemShut {Stop}%
\bibitem [{Not(){\natexlab{b}}}]{Not02}%
  \BibitemOpen
  \href@noop {} {} ({\natexlab{b}}),\ \bibinfo {note} {a simple model for VL
  disorder comprises the static Debye-Waller factor $W=\exp(-4\pi^2\langle u^2
  \rangle/a_0^2)$ where $\langle u^2 \rangle$ is the root mean square vortex
  displacement (along $\textbf{q}$) and $a_0=\sqrt{\sigma\Phi_0/H}$ is the
  vortex lattice spacing. Due to the similar field dependence of $\mathcal{F}$
  and $W$, it is not possible to disentangle contributions arising from the
  vortex core size $\xi$ and Debye-Waller effects.}\BibitemShut {Stop}%
\bibitem [{\citenamefont {Larkin}(1970)}]{Lar70}%
  \BibitemOpen
  \bibfield  {author} {\bibinfo {author} {\bibfnamefont {A.~I.}\ \bibnamefont
  {Larkin}},\ }\href@noop {} {\bibfield  {journal} {\bibinfo  {journal} {Sov.
  Phys. JETP},\ }\textbf {\bibinfo {volume} {31}},\ \bibinfo {pages} {784}
  (\bibinfo {year} {1970})}\BibitemShut {NoStop}%
\bibitem [{\citenamefont {Klein}\ \emph {et~al.}(2001)\citenamefont {Klein},
  \citenamefont {Joumard}, \citenamefont {Blanchard}, \citenamefont {Marcus},
  \citenamefont {Cubitt}, \citenamefont {Giamarchi},\ and\ \citenamefont
  {Le~Doussal}}]{Kle01}%
  \BibitemOpen
  \bibfield  {author} {\bibinfo {author} {\bibfnamefont {T.}~\bibnamefont
  {Klein}}, \bibinfo {author} {\bibfnamefont {I.}~\bibnamefont {Joumard}},
  \bibinfo {author} {\bibfnamefont {S.}~\bibnamefont {Blanchard}}, \bibinfo
  {author} {\bibfnamefont {J.}~\bibnamefont {Marcus}}, \bibinfo {author}
  {\bibfnamefont {R.}~\bibnamefont {Cubitt}}, \bibinfo {author} {\bibfnamefont
  {T.}~\bibnamefont {Giamarchi}}, \ and\ \bibinfo {author} {\bibfnamefont
  {P.}~\bibnamefont {Le~Doussal}},\ }\href@noop {} {\bibfield  {journal}
  {\bibinfo  {journal} {Nature},\ }\textbf {\bibinfo {volume} {413}},\ \bibinfo
  {pages} {404} (\bibinfo {year} {2001})}\BibitemShut {NoStop}%
\bibitem [{\citenamefont {Savici}\ \emph {et~al.}(2002)\citenamefont {Savici},
  \citenamefont {Fudamoto}, \citenamefont {Gat}, \citenamefont {Ito},
  \citenamefont {Larkin}, \citenamefont {Uemura}, \citenamefont {Luke},
  \citenamefont {Kojima}, \citenamefont {Lee}, \citenamefont {Kastner},
  \citenamefont {Birgeneau},\ and\ \citenamefont {Yamada}}]{Sav02}%
  \BibitemOpen
  \bibfield  {author} {\bibinfo {author} {\bibfnamefont {A.~T.}\ \bibnamefont
  {Savici}}, \bibinfo {author} {\bibfnamefont {Y.}~\bibnamefont {Fudamoto}},
  \bibinfo {author} {\bibfnamefont {I.~M.}\ \bibnamefont {Gat}}, \bibinfo
  {author} {\bibfnamefont {T.}~\bibnamefont {Ito}}, \bibinfo {author}
  {\bibfnamefont {M.~I.}\ \bibnamefont {Larkin}}, \bibinfo {author}
  {\bibfnamefont {Y.~J.}\ \bibnamefont {Uemura}}, \bibinfo {author}
  {\bibfnamefont {G.~M.}\ \bibnamefont {Luke}}, \bibinfo {author}
  {\bibfnamefont {K.~M.}\ \bibnamefont {Kojima}}, \bibinfo {author}
  {\bibfnamefont {Y.~S.}\ \bibnamefont {Lee}}, \bibinfo {author} {\bibfnamefont
  {M.~A.}\ \bibnamefont {Kastner}}, \bibinfo {author} {\bibfnamefont {R.~J.}\
  \bibnamefont {Birgeneau}}, \ and\ \bibinfo {author} {\bibfnamefont
  {K.}~\bibnamefont {Yamada}},\ }\href@noop {} {\bibfield  {journal} {\bibinfo
  {journal} {Phys. Rev. B},\ }\textbf {\bibinfo {volume} {66}},\ \bibinfo
  {pages} {014524} (\bibinfo {year} {2002})}\BibitemShut {NoStop}%
\bibitem [{\citenamefont {Uemura}(2003)}]{Uem03}%
  \BibitemOpen
  \bibfield  {author} {\bibinfo {author} {\bibfnamefont {Y.~J.}\ \bibnamefont
  {Uemura}},\ }\href@noop {} {\bibfield  {journal} {\bibinfo  {journal} {Solid
  State Comm.},\ }\textbf {\bibinfo {volume} {126}},\ \bibinfo {pages} {23}
  (\bibinfo {year} {2003})}\BibitemShut {NoStop}%
\bibitem [{\citenamefont {Dang}\ \emph {et~al.}(2010)\citenamefont {Dang},
  \citenamefont {Gull},\ and\ \citenamefont {Millis}}]{Dan10}%
  \BibitemOpen
  \bibfield  {author} {\bibinfo {author} {\bibfnamefont {H.~T.}\ \bibnamefont
  {Dang}}, \bibinfo {author} {\bibfnamefont {E.}~\bibnamefont {Gull}}, \ and\
  \bibinfo {author} {\bibfnamefont {A.~J.}\ \bibnamefont {Millis}},\
  }\href@noop {} {\bibfield  {journal} {\bibinfo  {journal} {Phys. Rev. B},\
  }\textbf {\bibinfo {volume} {81}},\ \bibinfo {pages} {235124} (\bibinfo
  {year} {2010})}\BibitemShut {NoStop}%
\bibitem [{\citenamefont {Panagopoulos}\ \emph {et~al.}(1999)\citenamefont
  {Panagopoulos}, \citenamefont {Rainford}, \citenamefont {Cooper},
  \citenamefont {Lo}, \citenamefont {Tallon}, \citenamefont {Loram},
  \citenamefont {Betouras}, \citenamefont {Wang},\ and\ \citenamefont
  {Chu}}]{Pan99}%
  \BibitemOpen
  \bibfield  {author} {\bibinfo {author} {\bibfnamefont {C.}~\bibnamefont
  {Panagopoulos}}, \bibinfo {author} {\bibfnamefont {B.~D.}\ \bibnamefont
  {Rainford}}, \bibinfo {author} {\bibfnamefont {J.~R.}\ \bibnamefont
  {Cooper}}, \bibinfo {author} {\bibfnamefont {W.}~\bibnamefont {Lo}}, \bibinfo
  {author} {\bibfnamefont {J.~L.}\ \bibnamefont {Tallon}}, \bibinfo {author}
  {\bibfnamefont {J.~W.}\ \bibnamefont {Loram}}, \bibinfo {author}
  {\bibfnamefont {J.}~\bibnamefont {Betouras}}, \bibinfo {author}
  {\bibfnamefont {Y.~S.}\ \bibnamefont {Wang}}, \ and\ \bibinfo {author}
  {\bibfnamefont {C.~W.}\ \bibnamefont {Chu}},\ }\href@noop {} {\bibfield
  {journal} {\bibinfo  {journal} {Phys. Rev. B},\ }\textbf {\bibinfo {volume}
  {60}},\ \bibinfo {pages} {14617} (\bibinfo {year} {1999})}\BibitemShut
  {NoStop}%
\end{thebibliography}%

\end{document}